\documentclass[twocolumn]{aastex631}
\usepackage{threeparttable}

\newcommand{\teff}{$T_{\rm eff}$}

\newcommand{\logg}{$\log g$}

\newcommand{\feh}{$\rm{[Fe/H]}$}

\newcommand{\vmic}{$v_{t}$}

\newcommand{\qq}{$\mathrm{q}^{2}$}

\newcommand{\sm}{$\rm{M_{\odot}}$}

\newcommand{\vsini}{$v \sin i$}

\newcommand{\be}{\begin{equation}}
\newcommand{\ee}{\end{equation}}
\newcommand{\ben}{\begin{eqnarray}}
\newcommand{\een}{\end{eqnarray}}
\newcommand{\bfg}{\begin{figure}}
\newcommand{\efg}{\end{figure}}

\newcommand{\gaiadrth}{\textit{Gaia} DR3}

\newcommand{\RHKteff}{$\log R^{'}_{\rm{HK}} (T_{\rm{eff}})$}
\newcommand{\li}{$A$(Li)}

\newcommand{\g}{$G$}

\newcommand{\Tc}{$T_{\rm{C}}$}

\shortauthors{Yana Galarza et al.}
\begin{document}

%\title{Template \aastex Article with Examples: 
%v6.3.1\footnote{Released on March, 1st, 2021}}

\title{HIP 8522: A Puzzling Young Solar Twin with the Lowest Detected Lithium\footnote{This research is based on data collected at the Subaru Telescope, which is operated by the National Astronomical Observatory of Japan (NAOJ). We are honored and grateful for the opportunity of observing the Universe from the Maunakea, which has cultural, historical, and natural significance in Hawaii. This paper includes data gathered with the 6.5 meter Magellan Telescopes located at Las Campanas Observatory (LCO), Chile.}}

\correspondingauthor{J. Yana Galarza}
\email{jyanagalarza@carnegiescience.edu}

\author[0000-0001-9261-8366]{Jhon Yana Galarza}
\altaffiliation{Carnegie Fellow}
\affiliation{The Observatories of the Carnegie Institution for Science, 813 Santa Barbara Street, Pasadena, CA 91101, USA}

\author[0000-0002-1387-2954]{Diego Lorenzo-Oliveira}
\affiliation{Laborat\'orio Nacional de Astrof\'isica, Rua Estados Unidos 154, 37504-364, Itajubá - MG, Brazil}

\author[0000-0003-2059-470X]{Thiago Ferreira}
\affiliation{Department of Astronomy, Yale University, 219 Prospect St., New Haven, CT 06511, USA}

\author[0000-0001-6533-6179]{Henrique Reggiani}
\affiliation{Gemini South, Gemini Observatory, NSF's NOIRLab, Casilla 603, La Serena, Chile}

\author[0000-0003-0012-9093]{Aida Behmard}
\affiliation{American Museum of Natural History, Dept. of Astrophysics, New York, NY, USA}

\author[0000-0002-4733-4994]{Joshua D. Simon}
\affiliation{The Observatories of the Carnegie Institution for Science, 813 Santa Barbara Street, Pasadena, CA 91101, USA}

\author[0000-0002-5084-168X]{Eder Martioli}
\affiliation{Laborat\'orio Nacional de Astrof\'isica, Rua Estados Unidos 154, 37504-364, Itajubá - MG, Brazil}

\author[0000-0002-7795-0018]{Ricardo López-Valdivia}
\affiliation{Universidad Nacional Autónoma de México, Instituto de Astronomía, AP 106, Ensenada 22800, BC, México}

\author[0000-0001-8179-1147]{Leandro de Almeida}
\affiliation{Laborat\'orio Nacional de Astrof\'isica, Rua Estados Unidos 154, 37504-364, Itajubá - MG, Brazil}

\author[0000-0002-8177-7633]{Emiliano Jofr\'e}
\affiliation{Universidad Nacional de C\'ordoba - Observatorio Astron\'{o}mico de C\'{o}rdoba, Laprida 854, X5000BGR, C\'ordoba, Argentina}
\affiliation{Consejo Nacional de Investigaciones Cient\'{i}ficas y T\'{e}cnicas (CONICET), Godoy Cruz 2290, CABA, CPC 1425FQB, Argentina}

\author[0000-0002-6871-1752]{Kareem El-Badry}
\affiliation{Department of Astronomy, California Institute of Technology, 1200 East California Boulevard, Pasadena, CA 91125, USA}

\begin{abstract}

We present HIP 8522, a young solar twin with the lowest detected lithium, potentially a field blue straggler or the result of episodic early accretion. Its stellar parameters (\teff\ $= 5729 \pm 7$ K, \logg\ $= 4.532 \pm 0.016$ dex, \feh\ $= 0.005 \pm 0.010$ dex, \vmic\ $= 1.08 \pm 0.02$ km s$^{-1}$) and chemical composition were determined via spectroscopic equilibrium using high resolution spectra ($R = 60~000-165~000$). The age of HIP 8522 was estimated to be an upper limit of $<$1 Gyr through isochrone fitting and was further confirmed using chemical clocks. Spectral synthesis of the lithium line at $\sim$6707.8 \AA\ yielded an upper lithium abundance limit of $A(\rm{Li}) <$ 0.8 dex. This value is unusually low for solar twins of similar age, which typically have $A(\rm{Li})$ values ranging from 2.0 to 3.3 dex, suggesting that $\sim$2 dex of lithium is missing. We investigate various scenarios, such as planet engulfment, sub-stellar mergers, and extra mixing. However, two distinct hypotheses provide plausible explanations for the significant depletion of lithium: one suggests that HIP 8522 is a field blue straggler formed by the merger of a close binary, while the other proposes that HIP 8522 experienced early episodic accretion. The young solar twin HIP 8522 presents an exceptional opportunity to rigorously test stellar evolution models and gain crucial insights into the internal mixing mechanisms responsible for the significant destruction of lithium.

\end{abstract}


\section{Introduction} 
\label{sec:intro}
Lithium ($^{7}$Li), beryllium ($^{9}$Be), and boron ($^{11}$B) are among the lightest elements in the Universe; destroyed below the convective zone of stars at temperatures of $\sim2.5\times 10^{6}$ K, $\sim3.5\times 10^{6}$ K, and $\sim5 \times 10^{6}$ K, respectively. Of these elements, lithium is the most sensitive to stellar evolution and the most studied, largely due to the easy access to its $\sim$6707.8 \AA\ line in the red part of the spectrum, despite significantly blending with CN molecules and other species. Besides, improvements in laboratory data allow for the high-precision measurement of lithium and facilitate exploration of topics such as planet engulfment in giant stars \citep{Aguilera:2016ApJ...833L..24A}, Galactic Chemical Evolution (GCE) of lithium \citep{Ramirez:2012ApJ...756...46R}, lithium evolution in open clusters \citep{Magrini:2021A&A...655A..23M, Aguilera:2023A&A...670A..73A}, the primordial lithium problem \citep{Coc:2009PhRvD..79j3512C, Kohri:2009PhRvD..79d3514K, Richard:2005ApJ...619..538R}, planet engulfment in solar twins\footnote{Solar twins are stars with stellar parameters within 100 K in \teff, and 0.1 dex in \logg\ and \feh\ of those of the Sun \citep{Ramirez:2014A&A...572A..48R}.} and binary systems \citep{Melendez:2017A&A...597A..34M, Saffe:2017A&A...604L...4S, Yana_Galarza:2021ApJ...922..129G, Liu:2024Natur.627..501L, Miquelarena:2024A&A...688A..73M}, and the constraint of interior stellar models \citep{Baraffe:2017A&A...597A..19B, Carlos:2016A&A...587A.100C, Martos:2023MNRAS.522.3217M, Rathsam:2023MNRAS.525.4642R}, among others.

 The study of lithium has sparked particular interest, as standard models of stellar interiors fail to account for the unexpectedly gradual depletion of lithium after the Zero Age Main Sequence (ZAMS) stage. These variations in Main Sequence (MS) stars were first reported by \citet{Herbig:1965ApJ...141..588H}, who suggested that lithium content could be used as an age indicator. Other studies have shown that lithium correlates with spectral type, \ion{Ca}{2} emission, and stellar rotation \citep[e.g.,][]{Wallerstein:1965ApJ...141..610W, Skumanich:1972ApJ...171..565S}. Following \citet{Herbig:1965ApJ...141..588H}, many early studies confirmed the age-dependent lithium depletion \citep[e.g.,][]{Conti:1966ApJ...146..383C, Danziger:1967ApJ...150..733D, Merchant:1966ApJ...143..336M, Wallerstein:1969ARA&A...7...99W, Zappala:1972ApJ...172...57Z}. Solar twins also follow the lithium-age correlation, and in the past decade and a half, the number of studies with large samples of solar twins has significantly increased \citep[e.g.,][]{Melendez:2010Ap&SS.328..193M, Monroe:2013ApJ...774L..32M, Carlos:2016A&A...587A.100C, Yana_Galarza:2016A&A...589A..17Y, Carlos:2019MNRAS.tmp..667C, Martos:2023MNRAS.522.3217M, Rathsam:2023MNRAS.525.4642R}. 
 
 To explain the lithium depletion in MS stars, non-standard models propose an additional mixing mechanism that transports lithium from the surface to below the convective zone, where it is destroyed. Examples of proposed extra mixing are rotation \citep{Pinsonneault:1989ApJ...338..424P, Chaboyer:1995ApJ...441..865C, Charbonnel:2005Sci...309.2189C}, microscopic diffusion and gravitational settling \citep{Michaud:2004ApJ...606..452M}, convective overshooting \citep{Xiong:2007ChA&A..31..244X, Baraffe:2017A&A...597A..19B}, internal gravity waves \citep{Charbonnel:2005Sci...309.2189C, Schlattl:1999A&A...347..272S, Baraffe:2010A&A...521A..44B}, and convective settling \citep{Andrassy:2015A&A...579A.122A}. Other sources of depletion include the removal of lithium from the atmosphere through mass loss \citep{Hobbs:1989ApJ...347..817H} and thermohaline convection \citep{Theado:2012ApJ...744..123T, Sevilla:2022MNRAS.516.3354S}. 

Since lithium can be destroyed by extra mixing mechanisms, it has been used as an indicator of planet engulfment events in solar twins, especially in binary systems with twin components \citep[e.g.,][]{Melendez:2017A&A...597A..34M, Saffe:2017A&A...604L...4S, Yana_Galarza:2021ApJ...922..129G, Emi_Jofre:2021AJ....162..291J, Spina:2021NatAs...5.1163S, Yong:2023MNRAS.526.2181Y, Flores:2024MNRAS.52710016F, Miquelarena:2024A&A...688A..73M, Yana:2024ApJ...974..122G, Liu:2024Natur.627..501L}. Although a significant enhancement of lithium is expected during a planet engulfment, simulations have shown that it quickly drops in a few billion years (depending on the mass of the star) through thermohaline convection and other mixing mechanisms \citep{Theado:2012ApJ...744..123T, Sevilla:2022MNRAS.516.3354S, Behmard:2023MNRAS.521.2969B}. This rapid depletion could erase the signatures of planet engulfment, making them difficult to detect. 

Another source of lithium depletion is related to planet formation. \citet{Delgado_Mena:2014A&A...562A..92D} suggested that exoplanets could potentially enhance mixing processes and significantly deepen the convective zone, leading to the burning of lithium. \citet{Martos:2023MNRAS.522.3217M}, using a sample of 118 solar-type stars, reported that planet-hosting stars show a depletion of lithium by $-0.23$ dex compared to their counterparts without planets (see their Figure 2). This finding supports the scenario where the formation of exoplanets can influence the lithium content of stars.

In an effort to increase the number of known solar twins, we created the Inti survey and reported the discovery of 40 new solar twins in the Northern Hemisphere \citep{Yana_Galarza:2021MNRAS.504.1873Y}. This new sample was analyzed using high-resolution spectra ($R = \lambda/\Delta \lambda = 60~000$) obtained with the Robert G. Tull Coud\'e Spectrograph (hereafter TS23; \citealp{Tull:1995PASP..107..251T}) mounted on the 2.7-m Harlan J. Smith Telescope at the McDonald Observatory. HIP 8522 (\gaiadrth\ 297548019938221056) is the second solar twin identified in our program, whose stellar parameters, age, mass and radius were reported in \citet{Yana_Galarza:2021MNRAS.504.1873Y}. This star was also analyzed by \citet{Notsu:2017PASJ...69...12N} as part of their study on the correlations between superflare frequencies, X-ray fluxes, and lithium abundances in solar-type stars. Both studies reported similar stellar parameters, but found a discrepant age. While they reported an age of 10 Gyr, we found a significantly younger age $<1$ Gyr (see Section \ref{sec:ages}). 

HIP 8522 stands out as a young solar twin, displaying chemical abundances, rotation, and activity levels in line with other young stars, but with one striking exception—its lithium abundance, which is surprisingly low ($<$0.8 dex). This unexpected finding forms the central focus of our study. Here, we present the discovery of this unusual low-lithium star and investigate the potential mechanisms responsible for its unexplained lithium depletion.

In Section \ref{sec:obs}, we describe the observations and data reduction. Section \ref{sec:stellaparam} presents the determination of stellar parameters, while Section \ref{sec:ages} covers the age estimation. Sections \ref{sec:chem} and \ref{sec:li} discuss the inferred chemical composition. Sections \ref{sec:kinematics} and \ref{sec:rv} describes the kinematics, spectral energy distribution, and radial velocities. In Section \ref{sec:disc}, we present the discussion of our results, and in section \ref{sec:conclusions} we provide a summary and conclusions. 

\begin{table}
\centering
\caption{Basics of the instruments for observing HIP 8522. The measured S/N is at $\sim$6500 \AA.}
\begin{tabular}{lccr}
\hline
\hline
 \bf{Instrument} &    \bf{$R$}       & \bf{S/N}       & \bf{Coverage}  \\
                 &    ($\lambda/\Delta \lambda$) & (pixel$^{-1}$) &     (\AA)     \\
 \hline
 TS23            & $60~000$          &       300      & $3700 - 9900$  \\
 HDS             & $165~000$         &       350      & $4400 - 7050$  \\
 MIKE            & $65~000 - 83~000$ &       180      & $3200 - 10000$ \\
 SOPHIE          & $75~000$          &       120      & $3872 - 6943$  \\
\hline
\end{tabular}
\label{tab:instruments}
\end{table}

\section{Observations and Reduction} \label{sec:obs}

\subsection{McDonald Observatory}
HIP 8522 was firstly observed with the TS23 spectrograph, which was configured in its high resolution mode ($R = 60~000$) on the 2.7 m Harlan J. Smith Telescope at the McDonald Observatory. We obtained two spectra of HIP 8522 and one spectrum of the Sun on 2019 August 25. Data reduction was performed with \textsc{PYRAF} \citep{PyRAF:2012ascl.soft07011S}, following the standard procedures that consist of bias subtraction, flat fielding, order extraction, and wavelength calibration. Then, the spectra were combined using the \textsc{scombine} task of \textsc{IRAF\footnote{\textsc{IRAF} is distributed by the National Optical Astronomical Observatories, which is operated by the Association of Universities for Research in Astronomy, Inc., under a cooperative agreement with the National Science Foundation.}}, resulting in a spectrum with signal-to-noise ratio (S/N pixel$^{-1}$) of $\sim$300 at $\sim$6500 \AA\ for HIP 8522, and 400 for the Sun, respectively. The TS23's spectral coverage ranges from $\sim 3700 - 9900$ \AA. 

\subsection{National Astronomical Observatory of Japan}
To confirm our initial results with TS23, we collected spectra of HIP 8522 with the High Dispersion Spectrograph \citep[HDS;][]{Noguchi:2002PASJ...54..855N}, Subaru programme (S22B-TE010-G), under the Gemini time exchange programme (BR\_2022B\_010), on the 8.2 m Subaru Telescope of the National Astronomical Observatory of Japan (NAOJ), located at the Mauna Kea summit. Data were taken on 2023 January 11. The observations were made using the image slicer IS \#3, with a slit of $0.2"\times×3$, providing a maximum resolving power of $R = 165~000$. We used the standard setup Yc (StdYc), which delivers a spectral coverage ranging from approximately 4400 \AA\ to 7050 \AA. Additionally, we observed the solar spectrum the same night through its light reflection on Jupiter's moon Ganymede, which is used as reference for our differential analysis (see Section \ref{sec:stellaparam}). We reduced the HDS/Subaru data using the Image Reduction and Analysis Facility (\textsc{IRAF}) framework\footnote{\url{https://www.naoj.org/Observing/Instruments/HDS/hdsql-e.html}}, performing flatfield and bias corrections, spectral order extractions, wavelength calibration, and removing the fringe pattern from the red side of the spectra. The coadded HDS spectra for HIP 8522 and the Sun have 350 and 450 of S/N at $\sim$6500 \AA, respectively.

\begin{table*}
	\caption{Comparison of the stellar parameters for HIP 8522.
$^{a}$ Trigonometric surface gravity was calculated following \citet{Yana_Galarza:2021MNRAS.504.1873Y}.
$^{b}$ Stellar parameters were adopted from \citet{Yana_Galarza:2021MNRAS.504.1873Y} using TS23. Isochronal ages and masses were calculated using each set of stellar parameters.}
	\label{tab:fundamental parameters}
	\centering
	\begin{tabular}{lccccccr} 
		\hline
		\hline
		\teff         & \logg             & \feh               & \vmic              & Age             & Mass              & \logg$^{a}$ & Instrument    \\
		(K)           & (dex)             & (dex)              &  (km s$^{-1}$)     & (Gyr)           & \sm               & (dex)             &           \\
		\hline 
		5727 $\pm$ 13 & 4.520 $\pm$ 0.026 & $-$0.017 $\pm$ 0.010 & 1.11 $\pm$ 0.03  & 0.60 $\pm$ 0.50 & 0.998 $\pm$ 0.008 & 4.543 $\pm$ 0.023 & TS23$^{b}$ ($R = 60~000$) \\
		5743 $\pm$ 10 & 4.570 $\pm$ 0.014 &   0.003 $\pm$ 0.010  & 1.06 $\pm$ 0.02  & 0.80 $\pm$ 0.45 & 1.013 $\pm$ 0.010 & 4.577 $\pm$ 0.022 & HDS ($R = 165~000$) \\
		5717 $\pm$ 15 & 4.505 $\pm$ 0.036 &   0.030 $\pm$ 0.013  & 1.08 $\pm$ 0.04  & $<0.43$ & 1.010 $\pm$ 0.010 & 4.531 $\pm$ 0.030 & MIKE ($R = 85~000$) \\
		\hline
		5729 $\pm$ 7 & 4.532 $\pm$ 0.016 &    0.005 $\pm$ 0.010 & 1.08 $\pm$ 0.02  & $<1$ & 1.007 $\pm$ 0.010 & 4.537 $\pm$ 0.020 & Adopted parameters \\
		\hline
		\hline
	\end{tabular}
\end{table*}

\begin{table}
    \begin{scriptsize}
    \centering
    \begin{threeparttable}
	\caption{Photometric and astrometric fundamental parameters for HIP 8522.} %The errors in \textit{Gaia} magnitudes were adopted from Vizier \citep{Ochsenbein:2000A&AS..143...23O}.}
    \begin{tabular}{lc}
    \hline 
	\hline
    {\bf Parameter} & {\bf Value}    \\
    \hline
    \multicolumn{2}{l}{\textbf{Photometric parameters}}                                  \\
    \textit{Gaia} DR3 $G$ [mag]                 & $8.401^{\alpha}$ \\%\pm0.003^{\alpha}$               \\
    \textit{Gaia} DR3 $(B-P)$ [mag]             & $8.732^{\alpha}$ \\% \pm0.003^{\alpha}$               \\
    \textit{Gaia} DR3 $(R-P)$ [mag]             & $7.893^{\alpha}$ \\%\pm0.004^{\alpha}$               \\
    \textit{Gaia} DR3 parallax [mas]            & $20.651\pm0.025^{\alpha}$              \\
    RUWE                                        & $0.987^{\alpha}$                       \\
    Radial velocity [km s$^{-1}$]               & $-2.17\pm0.22^{\alpha}$              \\
    2MASS $J$ [mag]                             & $7.33\pm0.02^{\beta}$                \\ 
    2MASS $H$ [mag]                             & $7.01\pm0.02^{\beta}$                \\ 
    2MASS $K_{\text{s}}$ [mag]                  & $6.94\pm0.02^{\beta}$                \\ 
    \rm{(B-V)} [mag]                                & $0.648\pm0.023^{\lambda}$               \\
    E(B-V) [mag]                                & $0.002\pm0.014^{\gamma}$               \\
    Distance [pc]                               & $48.331^{+0.053}_{-0.056}~^{\delta}$ \\
	\hline                                                                        
    \multicolumn{2}{l}{\textbf{Kinematics}}                        \\
    $z_{\rm{max}}$      [kpc]                   &  0.031                                 \\
    Eccentricity                                &  0.097                                \\
    Perigalacticon      [kpc]                   &  8.11                                 \\
    Apogalacticon       [kpc]                   &  9.84                                 \\
    $U$           [km s$^{-1}$]                 &  11.05                               \\
    $V$           [km s$^{-1}$]                 &  1.72                                \\
    $W$           [km s$^{-1}$]                 &  $-$6.46                              \\
    \hline                                                                        
    \multicolumn{2}{l}{\textbf{Rotation \& Activity parameters}}                        \\
    Period [days]                              & $7.7\pm 0.1^{\epsilon}$                \\
    v$\sin{i}$ [km s$^{-1}$]                   & $3.00\pm 0.16$                         \\
    $S_{\rm{MW}}$ (using TS23)                 & $0.324 \pm 0.003^{\epsilon}$           \\
    $S_{\rm{MW}}$ (using SOPHIE)               & $0.402 \pm 0.01$           \\
    \RHKteff                                   & $-4.53 \pm 0.08^{\epsilon}$            \\
    \hline
    \multicolumn{2}{l}{\textbf{Photometric \& Trigonometric inferred parameters}}       \\
    \teff~[K]                                  & $5714\pm 37^{\zeta}$                   \\
    \logg~[dex]                                & $4.54  \pm 0.02^{\epsilon}$          \\
    \hline                                                                        
    \multicolumn{2}{l}{\textbf{Ages from multiple methods}}                     \\
    Age from isochrones [Gyr]                  & $<1$                           \\
    Age from \RHKteff [Gyr]                    & $0.6^{+0.8}_{-0.4}~^{\eta}$    \\
    Age from Gyrochronology [Gyr]              & $<0.35^{\theta}$               \\
    Age from [Y/Mg] [Gyr]                      & $<0.3^{\iota}$                  \\
    Age from [Y/Al] [Gyr]                      & $<0.3^{\iota}$                  \\
    Age from photometric amplitudes [Gyr] & $<0.7^{\kappa}$ \\
    \hline
    \multicolumn{2}{l}{\textbf{Lithium abundance}}                     \\
    \li\ [dex]                  & $<0.8$ \\
    \hline                                                                        
    \end{tabular}
    \begin{tablenotes}
        \item $^{\alpha}$\cite{GaiaDR3:2023A&A...674A...1G}, $^{\beta}$\cite{2mass:2006AJ....131.1163S}, $^{\gamma}$\cite{Lallement:2014A&A...561A..91L, Lallement:2018A&A...616A.132L}, $^{\delta}$\cite{Bailer:2021AJ....161..147B}, $^{\epsilon}$\cite{Yana_Galarza:2021MNRAS.504.1873Y}, $^{\zeta}$\cite{Casagrande:2021MNRAS.507.2684C}, $^{\eta}$\cite{Diego:2018A&A...619A..73L}, $^{\theta}$\cite{Barnes:2010ApJ...722..222B}, $^{\iota}$\cite{Nissen:2017A&A...608A.112N},$^{\lambda}$\cite{Hipparcos_tycho:2000A&A...355L..27H}, $^\kappa$\citet{Ponte:2023MNRAS.522.2675P}.
    \end{tablenotes}
    \end{threeparttable}
    \label{tab:parameters}
    \end{scriptsize}
\end{table}

\subsection{Las Campanas Observatory}
We collected spectra of HIP 8522 with the Magellan Inamori Kyocera Echelle (MIKE) spectrograph on the Magellan Clay Telescope at Las Campanas Observatory \citep{Bernstein:2003SPIE.4841.1694B, Shectman:2003SPIE.4837..910S}. Data were taken on 2024 July 12. We used the $0.''35$ slit with standard blue and red grating azimuths, yielding spectra between 3200 and 10000 \AA\ with resolution of $83~000$ and $65~000$ in the blue and red arms, respectively. We also obtained the solar spectrum via the observation of the asteroid Vesta. The coadded MIKE spectra have S/Ns of 180 and 500 at $\sim$6500 \AA\ for HIP 8522 and the Sun, respectively.

\subsection{Observatoire de Haute-Provence}

We collected eight spectra of HIP 8522 from the online archive\footnote{\url{http://atlas.obs-hp.fr/sophie/}} of the Spectrographe pour l’Observation des Phénomènes des Intérieurs stellaires et des Exoplanètes (SOPHIE). SOPHIE is a stabilized high-resolution, fiber-fed, cross-dispersed échelle spectrograph mounted on the 1.93-m telescope at the Observatoire de Haute-Provence (OHP) \citep{Perruchot2008,Bouchy2013}. It covers a wavelength range from 3872 \AA\ to 6943 \AA\ across 39 spectral orders. We obtained the calibrated spectra, which were reduced using version 0.50 of the SOPHIE Data Reduction Software \citep[DRS,][]{Bouchy2009}. These spectra were observed in high spectral resolution mode (HR mode, $R=75~000$) in November 2007, January, August, and September 2008, and January 2016. The peak S/N per spectral element for these spectra ranged from approximately 15 to 70 (see Table \ref{tab:sophiequantities}).

The TS23, HDS and MIKE spectra were corrected by their barycentric velocities using \textsc{iSpec}\footnote{\url{https://www.blancocuaresma.com/s/iSpec}} \citep{Blanco:2014A&A...569A.111B, Blanco:2019MNRAS.486.2075B}. For the SOPHIE instrument, the radial velocities were estimated using the \textsc{sophie-toolkit} \citep{Martioli2023}, as discussed in Section \ref{sec:rv}. The rest frame spectra were individually normalized in pixel space with the \textsc{continuum} task in \textsc{IRAF}, by fitting the spectra with low degree spline functions (mostly third degree). We co-added the spectra, after normalization, with the \textsc{IRAF scombine} task using average values. The S/N and the spectral resolution $R$ achieved with each instrument are listed in Table \ref{tab:instruments}.

\section{Stellar Parameters} \label{sec:stellaparam}
The stellar parameters—effective temperature (\teff), surface gravity (\logg), metallicity (\feh), and microturbulent velocity (\vmic)—for HIP 8522 were previously reported by \citet{Yana_Galarza:2021MNRAS.504.1873Y}. With the acquisition of additional spectra at a higher resolution than TS23, using instruments such as HDS and MIKE, we have redetermined these stellar parameters to confirm our earlier results.

The stellar parameters of HIP 8522 were estimated through differential spectroscopic analysis relative to the Sun, based on measurements of the equivalent widths (EWs) of \ion{Fe}{1} and \ion{Fe}{2} atomic transitions. These measurements were obtained by fitting Gaussians to the line profiles, with pseudo-continuum regions of 6 \AA, using the \textsc{KAPTEYN} kmpfit package \citep{KapteynPackage} and the line list from \citet{Melendez:2014ApJ...791...14M}. It is important to highlight that this procedure was performed entirely manually, line by line, to ensure maximum precision in determining the stellar parameters. To prevent saturation effects, only iron lines with EW $<$ 130 m\AA\ were used for determining stellar parameters.

The \ion{Fe}{1} and \ion{Fe}{2} abundances were computed using the Qoyllur-quipu (\qq) Python code \citep{Ramirez:2014A&A...572A..48R}\footnote{\url{https://github.com/astroChasqui/q2_tutorial}}, which is configured to employ the Kurucz's \textsc{ODFNEW} model atmospheres \citep{Castelli:2003IAUS..210P.A20C} and the 2019 local thermodynamic equilibrium (LTE) code \textsc{moog} \citep{Sneden:1973PhDT.......180S} to reach the spectroscopic equilibrium. Table \ref{tab:fundamental parameters} lists the stellar parameters inferred from the HDS, and MIKE spectra, which are in good agreement with those estimated using TS23 within the uncertainties. We validated our spectroscopic \logg\ values using the trigonometric parallax method, as described in Equation (3) of \cite{Yana_Galarza:2021MNRAS.504.1873Y}, which includes \textit{Gaia} DR3 parallaxes, Johnson $V$ magnitudes \citep{Kharchenko:2001KFNT...17..409K}, bolometric corrections from \cite{Melendez:2006ApJ...641L.133M}, and stellar masses obtained through isochrone fitting (see Section \ref{sec:ages}). Table \ref{tab:fundamental parameters} shows no significant discrepancies between the trigonometric and spectroscopic surface gravity values for each set of stellar parameters. We also estimated photometric \teff\ using the Colour-\teff\ routine \textsc{colte}\footnote{\url{https://github.com/casaluca/colte}}, which uses \textit{Gaia} and 2MASS photometry in the InfraRed Flux Method (IRFM) to establish color-effective temperature relations \citep{Casagrande:2021MNRAS.507.2684C}. We found an average difference of 30 K between photometric and spectroscopic \teff\ using both \textit{Gaia} and 2MASS data, and 18 K using only \textit{Gaia}, which validates the excitation equilibrium. These additional analyses confirm the robustness of the determination of spectroscopic stellar parameters.

HIP 8522 is a young star ($<$1 Gyr; see Section \ref{sec:ages}), so it is expected to have a high activity level ($S_{\rm{MW}} \sim 0.32$ compared to the solar value of 0.17; see \citealp{Diego:2018A&A...619A..73L}). \citet{Yana:2019MNRAS.490L..86Y} demonstrated that the spectra of the young solar twin HIP 36515 are significantly affected by its magnetic activity, leading to variations in its stellar parameters throughout its six-year activity cycle. This could explain the different sets of stellar parameter values obtained for HIP 8522 (see Table \ref{tab:fundamental parameters}), which were measured at different epochs in its activity cycle. HIP 36515, like HIP 8522, is a young star ($<$1 Gyr) with an estimated activity cycle of six years. Although we lack sufficient data to estimate HIP 8522's activity cycle, its similarity in age and activity levels to HIP 36515 suggests a comparable activity cycle ($\sim$6 Gyr). \citet{Yana:2019MNRAS.490L..86Y} recommended adopting the stellar parameters measured at the minimum of the activity cycle; however, since the minimum of HIP 8522's cycle is unknown, we opted to use the average of all sets of stellar parameters, with uncertainties propagated quadratically, as listed in the last row of Table \ref{tab:fundamental parameters}. \citet{Notsu:2017PASJ...69...12N} estimated similar stellar parameters (\teff\ = 5705 $\pm$ 15 K, \logg\ = 4.59 $\pm$ 0.04 dex, \feh\ $= -0.04 \pm 0.02$ dex, \vmic\ = 1.01 $\pm$ 0.08 km s$^{-1}$), which are in agreement with our results. Table \ref{tab:parameters} lists the photometric effective temperature, trigonometric \logg, ages from chemical clocks, photometric data, and other relevant parameters for HIP 8522.
 
\section{Stellar ages} \label{sec:ages}
We estimated age and mass using the Bayesian method implemented in \qq, which employs the Yonsei–Yale isochrones of stellar evolution \citep{Yi:2001ApJS..136..417Y, Demarque:2004ApJS..155..667D} with the adopted stellar parameters, \textit{Gaia} DR3 $G$ magnitude and parallax as input parameters, strictly following the procedure given in \cite{Yana_Galarza:2021MNRAS.504.1873Y}. The results are displayed in Table \ref{tab:fundamental parameters}. All the sets of stellar parameters yield similar isochronal age within the uncertainties, confirming that HIP 8522 is a young solar twin ($<$1 Gyr). 

We conducted further analysis using age indicators established from solar twins to confirm the isochronal ages. Using the activity-age relation from \citet[][Eq. 11]{Diego:2018A&A...619A..73L}, we estimated an age of $0.6^{+0.8}_{-0.4}$ Gyr, which is close to the youngest group of activity calibrators defined by the authors ($t_{\rm HK, T_{eff}} = 0.60^{+0.19}_{-0.14}$ Gyr, see their Sec. 3.2 and Fig. 7). In Section \ref{sec:chem}, we reported the chemical abundance of HIP 8522 and found [Y/Mg] and [Y/Al] values of 0.141 dex and 0.183 dex, respectively. These values correspond to an upper limit on the age of less than 0.30 Gyr in the [Y/Mg]-age and [Y/Al]-age calibrations of \citet[][see Equations (1) and (2) therein]{Nissen:2017A&A...608A.112N}. The rotational period, estimated to be 7.7 days in \citet{Yana_Galarza:2021MNRAS.504.1873Y}, yields an age of $<0.35$ Gyr using the period evolution equation from \citet{Barnes:2010ApJ...722..222B}, which matches the isochronal ages. \citet{Ponte:2023MNRAS.522.2675P} reported that TESS photometric amplitudes are correlated with stellar ages, providing an independent age estimator. Following their prescription in Section 3, we estimated a logarithm amplitude of 3.8 [erg/cm$^{2}$/s], corresponding to an upper limit on the age of $<$0.7 Gyr using their Equation 8. The $v \sin{i}$ estimated in 3.0 $\pm$ 0.2 km s$^{-1}$ (see Section \ref{sec:li}) suggests that HIP 8522's age is lower than 1 Gyr \citep[][see their Figure 5]{Leo:2016A&A...592A.156D}. Together, these findings support that HIP 8522 is a young solar twin. The ages derived from these correlations are summarized in Table \ref{tab:parameters}. Another straightforward example supporting our conclusion is the comparison of its kinematics with those of stars of similar and older ages. The upper panel of Fig. \ref{fig:toomre_sed} depicts the Toomre diagram, showing the space velocities for HIP 8522 (see Section \ref{sec:kinematics}). It shows that HIP 8522 is part of the group of young stars ($\leq1$ Gyr; yellow crosses) and is distant from the group of older stars ($\geq7$ Gyr; blue squares).

As mentioned earlier, \citet{Notsu:2017PASJ...69...12N} estimated the age of HIP 8522 to be 10 Gyr using tracks from PARSEC isochrones \citep{Bressan:2012MNRAS.427..127B}, which is significantly different from our estimate. We performed a simple experiment: we used Notsu's stellar parameters and ran \qq, first using only the spectroscopic surface gravity and then including the parallax. We found ages of $1.3 \pm 1.1$ Gyr and $0.6 \pm 0.5$ Gyr, which are consistent with our result within the uncertainties. The difference between our method and that of \citet{Notsu:2017PASJ...69...12N} is that they used only three input parameters—luminosity, metallicity, and effective temperature (see their Appendix 1.8)—which could affect the probability distribution functions for age and result in an inaccurate age. Additionally, they use an interstellar extinction value of $A_{V}$ = 0.1, which is significantly different from the value we adopted, $A_{V}$ = 0.0062 (estimated assuming a visual extinction-to-reddening ratio of $A_{V}$/E(B$-$V) = 3.1 \citep{ODonnell:1994ApJ...422..158O}, with E(B$-$V) = 0.002 adopted from \citealp{Lallement:2014A&A...561A..91L, Lallement:2018A&A...616A.132L}). To explore whether the difference was due to different evolutionary models, we used evolutionary tracks from the Dartmouth isochrones \citep{Dotter:2008ApJS..178...89D} and consistently found that, regardless of the set of stellar parameters, HIP 8522's age is $<1$ Gyr. On the other hand, \citet{Notsu:2017PASJ...69...12N} reported the ages of 18 Sco and HIP 100963 to be $6.8 \pm 0.2$ Gyr and $9.3 \pm 0.7$ Gyr, respectively (see their Table 3). These stars are solar twins, with their ages estimated in several other studies. For instance, for 18 Sco, \citet{Nissen:2020A&A...640A..81N} reported an age of 3.9 $\pm$ 0.7 Gyr, while \citet{Melendez:2014ApJ...791...14M} found 2.1 $\pm$ 1.0 Gyr, and \citet{Spina:2018MNRAS.474.2580S} reported $3.2 \pm 0.9$ Gyr. For HIP 100963, \citet{Takeda:2009PASJ...61..471T} estimated an age of 4.6 Gyr, and \citet{Boesgaard:2022ApJ...941...21B} reported 5.5 Gyr. These results differ from the ages estimated by \citet{Notsu:2017PASJ...69...12N} for both solar twins.

\begin{table*}
    \centering
	\caption{$^{(\alpha)}$Differential chemical abundances for HIP 8522 relative to the Sun. $^{(\beta)}$GCE-corrected abundances. $^{(\gamma)}$Predicted abundances from a model of planet engulfment \citep{Behmard:2023MNRAS.521.2969B}. The last column is the 50\% dust condensation temperature of elements \citep{Lodders:2003ApJ...591.1220L}. $^{(\delta)}$ Weighted average of \ion{Fe}{1} and \ion{Fe}{2}. $^{(\omega)}$Absolute abundance of lithium \li.}
	%\begin{scriptsize}
	\begin{tabular}{lccccr} 
		\hline
		\hline
		Element            &   Z   & HIP 8522$^{(\alpha)}$  & HIP 8522$^{(\beta)}$     &     Model$^{(\gamma)}$                        & $T_{\rm{Cond}}$  \\   
		                   &       &    $\Delta$[X/H] (dex)  &    $\Delta$[X/H] (dex) &  $\Delta$[X/H]$_{\rm{(Star-Sun)}}$ (dex)  &      (K)         \\   
        \hline                                                                                                                                                                                              
		Li$^{(\omega)}$    &  3    &  $<$0.8   &      \ldots                 &        \ldots                        & 1142 \\
        \ion{C}{1}         &  6    &  $-$0.070 $\pm$ 0.020   &  $-$0.022 $\pm$ 0.022       &        $-$0.027                      & 40 \\
        \ion{O}{1}         &  8    &  $-$0.021 $\pm$ 0.019   &  $+$0.016 $\pm$ 0.020       &        $-$0.021                      & 180 \\
        \ion{Na}{1}        &  11   &  $-$0.084 $\pm$ 0.009   &  $-$0.048 $\pm$ 0.012       &        $-$0.020                      & 958 \\
        \ion{Mg}{1}        &  12   &  $-$0.045 $\pm$ 0.014   &  $-$0.004 $\pm$ 0.015       &        $-$0.004                      & 1336 \\
        \ion{Al}{1}        &  13   &  $-$0.087 $\pm$ 0.015   &  $-$0.029 $\pm$ 0.017       &        $+$0.003                      & 1653 \\
        \ion{Si}{1}        &  14   &  $-$0.031 $\pm$ 0.005   &  $-$0.005 $\pm$ 0.006       &        $-$0.001                      & 1310 \\
        \ion{S}{1}         &  16   &  $-$0.016 $\pm$ 0.016   &  $+$0.025 $\pm$ 0.018       &        $-$0.025                      & 664 \\
        \ion{K}{1}         &  19   &  $-$0.069 $\pm$ 0.022   &  $-$0.069 $\pm$ 0.022       &        $-$0.021                      & 1006 \\
        \ion{Ca}{1}        &  20   &  $+$0.023 $\pm$ 0.013   &  $+$0.018 $\pm$ 0.013       &        $+$0.001                      & 1517 \\
        \ion{Sc}{1}        &  21   &  $-$0.017 $\pm$ 0.014   &  $-$0.003 $\pm$ 0.012       &        $-$0.002                      & 1659 \\
        \ion{Sc}{2}        &  21   &  $-$0.050 $\pm$ 0.021   &  $-$0.026 $\pm$ 0.012       &        $-$0.002                      & 1659 \\
        \ion{Ti}{1}        &  22   &  $-$0.001 $\pm$ 0.014   &  $+$0.005 $\pm$ 0.010       &        $+$0.001                      & 1582 \\
        \ion{Ti}{2}        &  22   &  $-$0.020 $\pm$ 0.015   &  $-$0.005 $\pm$ 0.010       &        $+$0.001                      & 1582 \\
        \ion{V}{1}         &  23   &  $+$0.018 $\pm$ 0.018   &  $+$0.023 $\pm$ 0.018       &        $+$0.008                      & 1429 \\
        \ion{Cr}{1}        &  24   &  $+$0.017 $\pm$ 0.014   &  $+$0.005 $\pm$ 0.010       &        $+$0.003                      & 1296 \\
        \ion{Cr}{2}        &  24   &  $+$0.006 $\pm$ 0.015   &  $-$0.001 $\pm$ 0.010       &        $+$0.003                      & 1296 \\
        \ion{Mn}{1}        &  25   &  $-$0.036 $\pm$ 0.012   &  $-$0.026 $\pm$ 0.013       &        $-$0.019                      & 1158 \\
        Fe$^{(\delta)}$    &  26   &  $+$0.002 $\pm$ 0.010   &  $+$0.002 $\pm$ 0.010       &        $-$0.001                      & 1334 \\
        \ion{Co}{1}        &  27   &  $-$0.049 $\pm$ 0.015   &  $-$0.018 $\pm$ 0.016       &        $-$0.005                      & 1352 \\
        \ion{Ni}{1}        &  28   &  $-$0.068 $\pm$ 0.010   &  $-$0.038 $\pm$ 0.011       &        $+$0.000                      & 1353 \\
        \ion{Cu}{1}        &  29   &  $-$0.084 $\pm$ 0.024   &  $-$0.022 $\pm$ 0.026       &        $-$0.018                      & 1037 \\
        \ion{Zn}{1}        &  30   &  $-$0.083 $\pm$ 0.016   &  $-$0.040 $\pm$ 0.018       &        $-$0.025                      & 726 \\
        \ion{Sr}{1}        &  38   &  $+$0.114 $\pm$ 0.027   &  $+$0.009 $\pm$ 0.032       &        \ldots                        & 1464 \\
        \ion{Y}{2}         &  39   &  $+$0.096 $\pm$ 0.015   &  $-$0.004 $\pm$ 0.022       &        \ldots                        & 1659 \\
        \ion{Ba}{2}        &  56   &  $+$0.189 $\pm$ 0.011   &  $+$0.056 $\pm$ 0.021       &        \ldots                        & 1455 \\
        \ion{La}{2}        &  57   &  $+$0.139 $\pm$ 0.019   &  $+$0.044 $\pm$ 0.024       &        \ldots                        & 1578 \\
        \ion{Ce}{2}        &  58   &  $+$0.128 $\pm$ 0.017   &  $+$0.036 $\pm$ 0.022       &        \ldots                        & 1478 \\
        \ion{Nd}{2}        &  60   &  $+$0.128 $\pm$ 0.017   &  $+$0.045 $\pm$ 0.021       &        \ldots                        & 1602 \\
        \ion{Sm}{2}        &  62   &  $+$0.094 $\pm$ 0.026   &  $+$0.062 $\pm$ 0.027       &        \ldots                        & 1590 \\
        \ion{Eu}{2}        &  63   &  $+$0.066 $\pm$ 0.027   &  $+$0.043 $\pm$ 0.028       &        \ldots                        & 1356 \\
		\hline
		\hline
	\end{tabular}
	%\end{scriptsize}
	\label{tab:chemical_abundances}
\end{table*}

\section{Chemical Composition} \label{sec:chem}
The chemical abundance of HIP 8522 was estimated by measuring equivalent widths (EWs) using a procedure similar to that used for \ion{Fe}{1} and \ion{Fe}{2}, following the differential technique implemented in \qq. Since the HDS spectrum has the highest resolution and S/N, we used it to calculate the chemical composition. However, its short spectral coverage does not include elements like oxygen, potassium, and aluminum. Those elements were measured using the TS23 and MIKE spectra, which have a larger wavelength coverage than HDS (see Table \ref{tab:instruments}). %We did not find significant differences between the abundances from the different instruments; they agree on average within 0.02 dex.

In total, we obtained abundances for 31 species: \ion{C}{1}, \ion{O}{1}. \ion{Na}{1}, \ion{Mg}{1}, \ion{Al}{1}, \ion{Si}{1}, \ion{S}{1}, \ion{Ca}{1}, \ion{Sc}{1}, \ion{Sc}{2}, \ion{Ti}{1}, \ion{Ti}{2}, \ion{V}{1}, \ion{Cr}{1}, \ion{Cr}{2}, \ion{Mn}{1}, \ion{Fe}{1}, \ion{Fe}{2}, \ion{Co}{1}, \ion{Ni}{1}, \ion{Cu}{1}, \ion{Zn}{1}, \ion{Sr}{1}, \ion{Y}{2}, \ion{Ba}{2}, \ion{La}{2}, \ion{Ce}{2}, \ion{Nd}{2}, \ion{Sm}{2}, \ion{Eu}{2}, and \ion{Li}{1}. Oxygen abundances were inferred from the high-excitation \ion{O}{1} $\lambda 777$ nm triplet using TS23/MIKE spectra and corrected for NLTE employing the grids of \citet{Ramirez:2007A&A...465..271R}. We accounted for hyperfine structure and isotopic splitting for \ion{Sc}{1}, \ion{Sc}{2}, \ion{V}{1}, \ion{Mn}{1}, \ion{Co}{1}, \ion{Cu}{1}, \ion{Y}{2}, \ion{Ba}{2}, and \ion{Eu}{2} from \cite{McWilliam:1998AJ....115.1640M,Prochaska2000a,Prochaska2000b,klose2002,Cohen:2003ApJ...588.1082C,Blackwell-Whitehead2005a,Blackwell-Whitehead2005b,Lawler2014}, and from the Kurucz\footnote{\url{http://kurucz.harvard.edu/linelists.html}} line lists. The chemical abundances of HIP 8522 relative to the Sun are summarized in Table \ref{tab:chemical_abundances}.

\begin{figure}
 \includegraphics[width=\columnwidth]{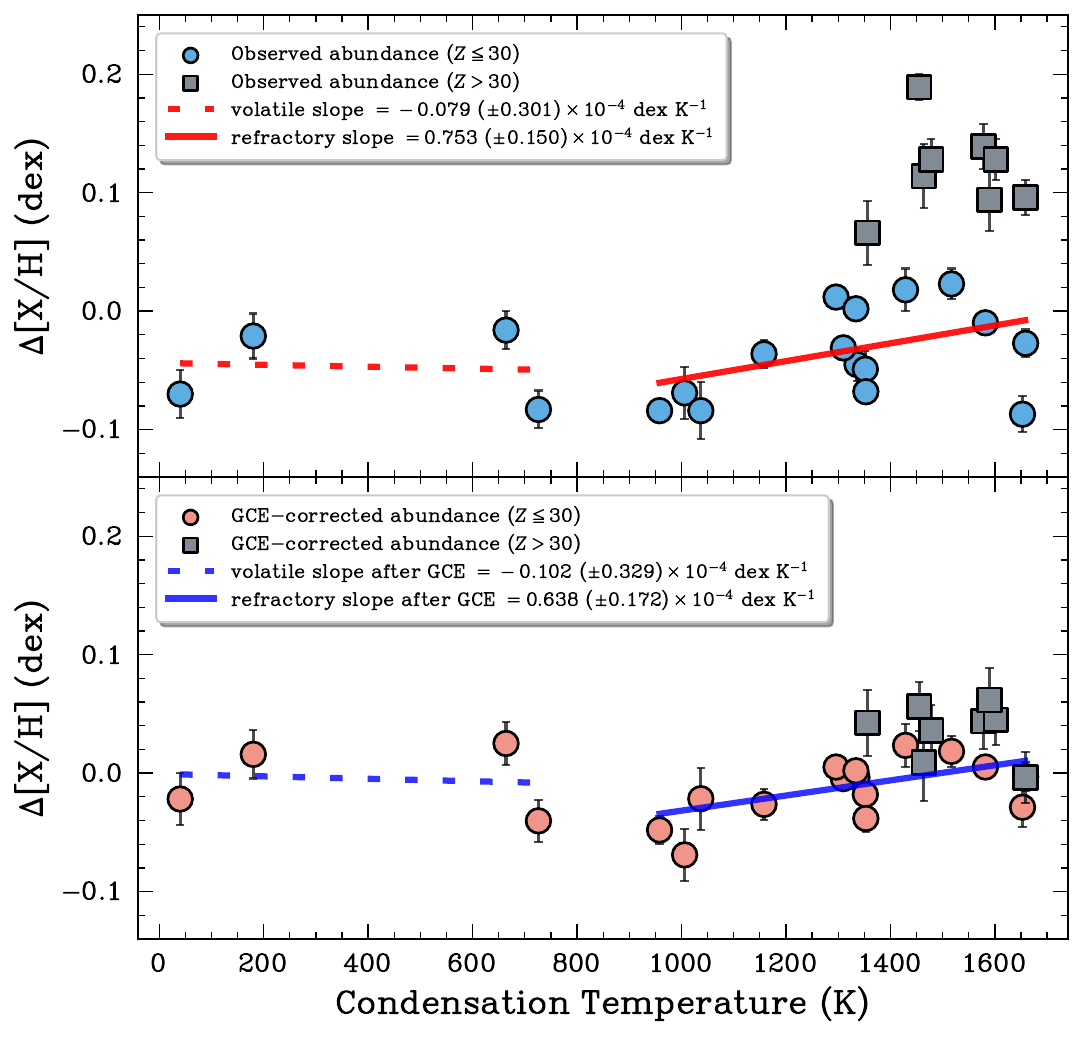}
 \centering
 \caption{Observed (upper panel) and GCE-corrected (bottom panel) differential chemical abundances for HIP 8522 relative to the Sun as a function of dust condensation temperature \citep{Lodders:2003ApJ...591.1220L}. The solid and dashed lines are the fits to the refractory (\Tc\ $>$ 900 K) and volatile elements (\Tc\ $<$ 900 K), respectively.}
 \label{fig:tc_abundances}
\end{figure}

The upper panel of Figure \ref{fig:tc_abundances} shows the differential chemical abundance for HIP 8522 relative to the Sun ($\Delta$[X/H]) as a function of condensation temperature\footnote{Temperature at which 50\% of an element condenses from a gas with solar composition at a total pressure of $10^{-4}$ bar \citep{Lodders:2003ApJ...591.1220L}.} (\Tc). This figure clearly shows the division of elements into three groups: volatile elements with \Tc\ lower than 900 K, refractory elements with \Tc\ higher than 900 K, and neutron-capture elements (gray squares). It also shows that HIP 8522 is slightly enhanced in refractory elements (blue circles with \Tc $>$ 900 K) but significantly enhanced in neutron capture elements. Nevertheless, to properly compare HIP 8522 to the Sun and detect chemical anomalies, it is necessary to account for the Galactic Chemical Evolution (GCE) of stars and place the star on the same age scale as the Sun. The correction is performed using the correlations between individual abundances and ages, provided by \cite{Nissen:2015A&A...579A..52N, Bedell:2018ApJ...865...68B, Spina:2018MNRAS.474.2580S}, based on solar twins, accounting for the effects of metallicity. Consistent with our previous studies on other solar twins, we applied the corrections from \cite{Bedell:2018ApJ...865...68B}, following Equation (1) in \citet{Yana_Galarza:2016A&A...589A..17Y}. The bottom panel of Figure \ref{fig:tc_abundances} shows the GCE-corrected abundances as a function of \Tc, where the refractory elements remain enhanced compared to the Sun. This is shown by the linear fit (solid blue line) to those elements, with a slope statistically significant at $\sim 4\sigma$. In subsection \ref{sub:engulfment}, we discuss in detail the implications of the chemical anomalies observed in HIP 8522.

\begin{figure}
 \includegraphics[width=\columnwidth]{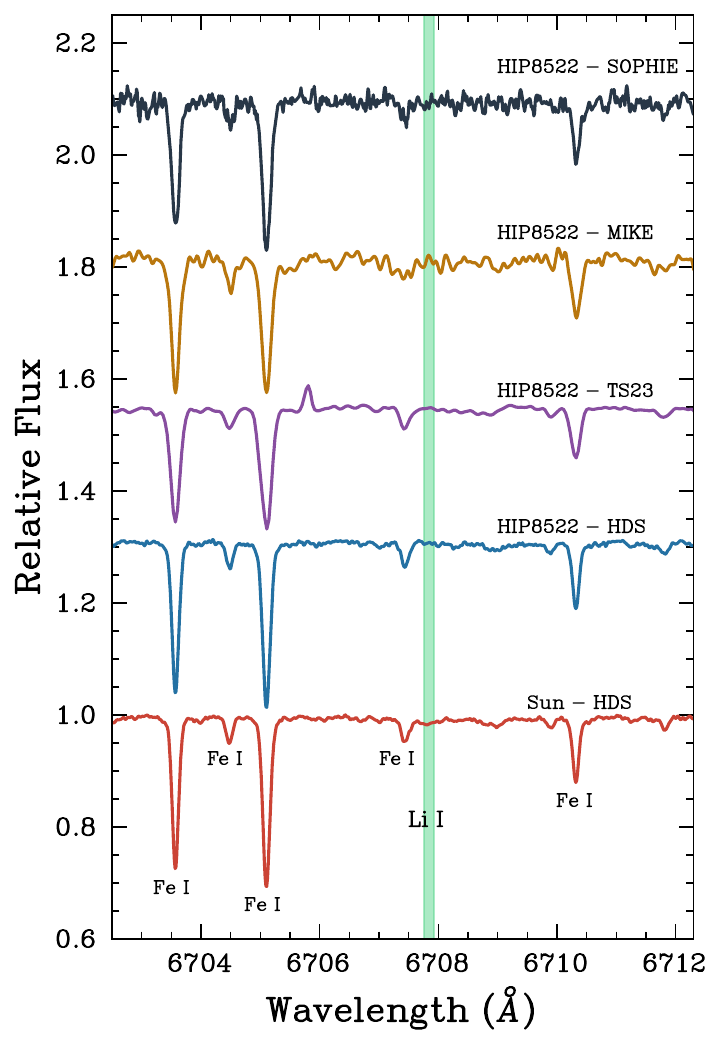}
 \centering
 \caption{SOPHIE ($R = 75~000$), MIKE ($R = 65~000$), TS23 ($R = 60~000$), and HDS ($R = 165~000$) spectra for HIP 8522 (from top to bottom). The lithium line is highlighted by the green-shaded region. The red spectrum represents the Sun, obtained with HDS.}
 \label{fig:Li_all_instruments}
\end{figure}

\section{Lithium abundance} \label{sec:li}
We determined the lithium abundance (hereafter \li) following the procedure outlined in \citet{Yana_Galarza:2016A&A...589A..17Y}, which involves measuring line broadening and performing spectral synthesis of the lithium feature at $\sim$6707.8 \AA. We used the HDS spectrum because it has the highest resolution and S/N compared to TS23, MIKE, and SOPHIE (see Figure \ref{fig:Li_all_instruments}). The procedure consists, firstly, of determining the solar macroturbulent velocity ($v_{\rm{macro, \odot}}$) using the HDS solar spectrum, as this value depends on the spectral resolution and may not necessarily match what is reported in other studies. Using inaccurate $v_{\rm{macro, \odot}}$ values could introduce systematic errors in the lithium measurements, particularly when the Sun is used as the reference star. We fixed \vsini$_{\odot}$ at 1.9 km s$^{-1}$ \citep{Bruning:1984ApJ...281..830B, Saar:1997MNRAS.284..803S} and through spectral synthesis of five iron lines (6027.050 \AA, 6093.644 \AA, 6151.618 \AA, 6165.360 \AA, 6705.102 \AA) and one of nickel (6767.772 \AA), we obtained consistent macroturbulent velocities for each line, with an average of  $v_{\rm{macro, \odot}} = 3.30 \pm 0.15$ km s$^{-1}$. This is in good agreement with the values reported by \citet[][see their Table 2]{Doyle:2014MNRAS.444.3592D}, which were obtained using solar spectra with resolutions ranging from $R=76~000-300~000$. The spectral synthesis was performed using \textsc{MOOG} and Kurucz \textsc{ODFNEW} model atmospheres, adopting the standard solar stellar parameters (\teff\ = 5777 K, \logg\ = 4.44 dex, \feh\ = 0.0 dex, \vmic\ = 1.0 km s$^{-1}$). 

\begin{figure}
 \includegraphics[width=\columnwidth]{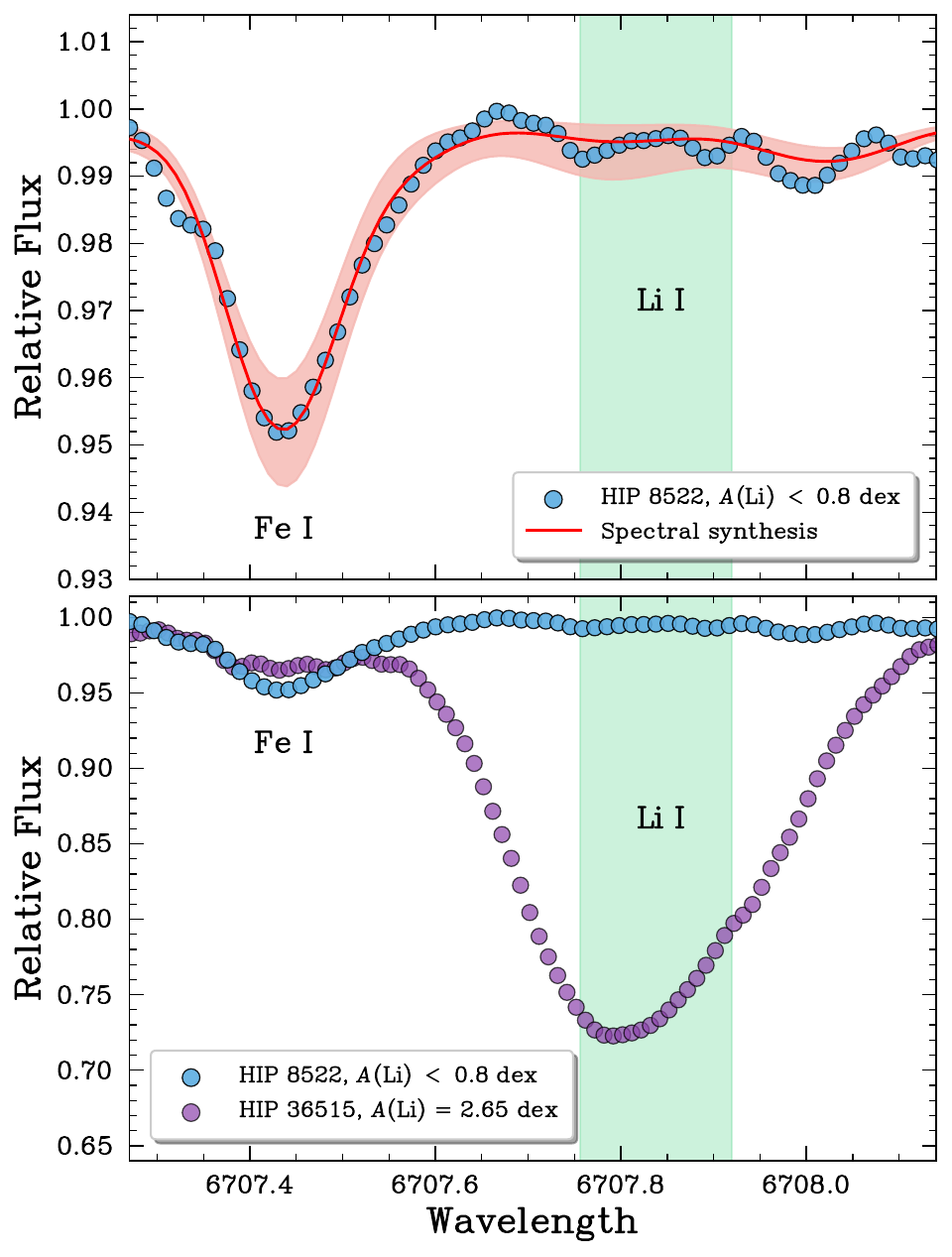}
 \centering
 \caption{\textbf{Upper panel:} Observed spectra of HIP 8522 (blue circles) around the lithium doublet at $\sim$6708 \AA. The spectral synthesis is shown by the red solid line. The red shaded region represents a 0.1 dex uncertainty budget, except for lithium, which has an upper limit of $<0.8$ dex. \textbf{Bottom panel:} Comparison between HIP 8522 (blue circles) and HIP 36515 (purple circles; \citealp{Yana:2019MNRAS.490L..86Y}). Both stars are solar twins with similar stellar parameters, including age, activity level, and rotational period, but their lithium abundances differ by $\sim$1.8 dex. The lithium line is highlighted by the green-shaded region in both panels.}
 \label{fig:Li_fit}
\end{figure}

Once $v_{\rm{macro, \odot}}$ is calculated, we use this value in Equation (1) from \citet{Leo:2016A&A...592A.156D} to determine the macroturbulence velocities of HIP 8522. By synthesizing the iron and nickel lines, we then determine the projected rotational velocity, obtaining an average $v_{\rm{macro}} = 2.95 \pm 0.16$ km s$^{-1}$ and \vsini\ = $3.00 \pm 0.16$ km s$^{-1}$ for HIP 8522. \citet{Notsu:2017PASJ...69...12N} reported a similar \vsini\ (2.6 $\pm$ 0.4 km s$^{-1}$). 

Next, we performed spectral synthesis of the Li line using the line list from \citet{Melendez:2012A&A...543A..29M}, the estimated broadening velocities, and the \textsc{synth} driver in MOOG. This resulted in an LTE abundance upper limit of $<$0.8 dex, as shown in the upper panel of Fig. \ref{fig:Li_fit}. This value was then corrected for NLTE effects using the data from \citet{Lind:2009A&A...503..541L}\footnote{\url{www.inspect-stars.com}}, yielding \li$_{\rm{NLTE}} < 0.794$ dex. We repeated the same procedure to estimate lithium in the Sun (\li$_{\rm{\odot, LTE}} = 1.05$ dex) and obtained \li$_{\rm{\odot, NLTE}} = 1.08 \pm 0.03$ dex, which is consistent within the errors with the value obtained by \citet[][\li\ $= 1.05 \pm 0.1$ dex]{Asplund:2009ARA&A..47..481A} using a three-dimensional hydrodynamical model of the solar atmosphere. 

HIP 8522 shows unexpectedly high lithium depletion compared to solar twins of the same age (see upper panel of Fig. \ref{fig:unique}). In the bottom panel of Fig. \ref{fig:Li_fit}, we compare the spectra of HIP 8522 with HIP 36515, a solar twin with similar stellar parameters (\teff\ = 5843 $\pm$ 10 K, \logg\ = 4.52 $\pm$ 0.02 dex, \feh\ $= -0.032 \pm 0.01$ dex), age ($0.7 \pm 0.5$ Gyr), stellar activity level ($S_{\rm{MW}} = 0.301 \pm 0.004$), and rotation period (P$_{\rm rot}$ = $4.6 \pm 1.2$ d). For more details we refer to \citet{Yana:2019MNRAS.490L..86Y}. However, the lithium abundance differs by $\sim1.8$ dex between the two stars. In Section \ref{sec:disc}, we will discuss the implications of this rare finding.

\section{Kinematics and Spectral Energy Distribution} \label{sec:kinematics}

Given the significant depletion of lithium observed in HIP 8522, in this section we analyze its membership in the Galactic disc, as well as investigate the possibility of an unresolved sub-stellar companion.

\begin{figure}
    $\begin{array}{c}
    \includegraphics[width=\columnwidth]{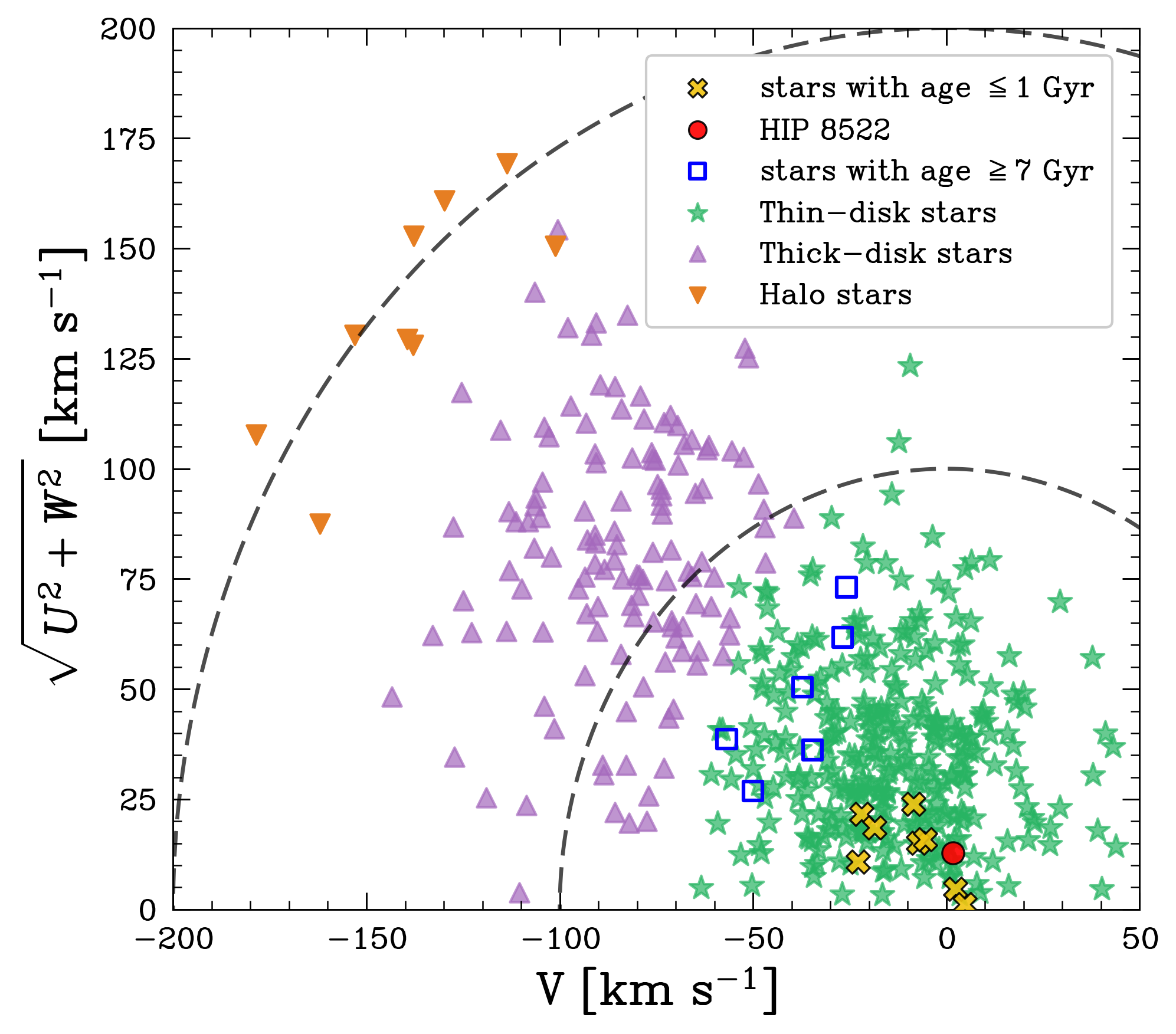} \\
    \includegraphics[width=\columnwidth]{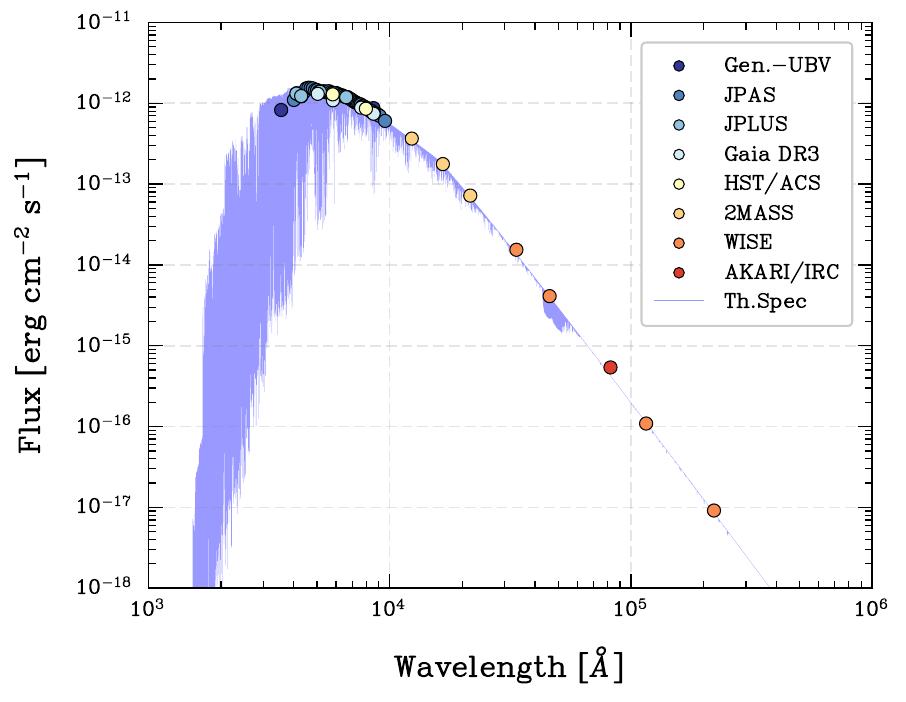}
    \end{array}$
    \centering
    \caption{\textbf{Upper panel:} Toomre diagram for HIP 8522 (red circle). The space velocities, calculated with {\sc Gala}, confirm that the star belongs to the thin-disc. For comparison, stars, triangles, and inverted triangles represent stars from the thin-disc, thick-disc, and halo of the Galaxy, respectively. Data taken from \citet{Ramirez:2007A&A...465..271R}. The yellow crosses represent solar twins with ages $\leq 1$ Gyr, while the blue squares represent solar twins with ages $\geq$7 Gyr. \textbf{Bottom panel:} SED fitting for HIP 8522. The colored circles correspond to the observed photometric data, while the purple solid line is the best-fit synthetic spectra.}
    \label{fig:toomre_sed}
\end{figure}

We used the \textsc{Gala}\footnote{\url{https://github.com/adrn/gala}} code \citep{gala, adrian_price_whelan_2020_4159870} along with \textit{Gaia} DR3 proper motions and radial velocities \citep{GaiaDR3:2023A&A...674A...1G} to calculate the Galactic orbits of HIP 8522. We configured \textsc{Gala} to use the default {\sc MilkyWayPotential} model, which includes a spherical nucleus and bulge \citep{Hernquist:1990ApJ...356..359H}, a Miyamoto–Nagai disk \citep{Miyamoto:1975PASJ...27..533M, Bovy:2015ApJS..216...29B}, and a spherical Navarro-Frenk-White (NFW) dark matter halo \citep{Navarro:1996ApJ...462..563N}. We adopted the Sun's position and velocity as $x_{\odot} = (-8.3, 0, 0)$ kpc and $v_{\odot} = (-11.1, 244, 7.25)$ km s$^{-1}$ \citep{Schonrich:2010MNRAS.403.1829S, Schonrich:2012MNRAS.427..274S}. We performed orbital integration with a 0.5 Myr timestep over 2 Gyr. The Galactic space velocities ($U, V, W$, see Table \ref{tab:parameters}) indicate that HIP 8522 is located right within the thin-disc kinematic distribution on the Toomre diagram  ($V$ vs. $\sqrt{U^{2}+W^{2}}$; see upper panel of Figure \ref{fig:toomre_sed}). This result is supported by the [$\alpha$/Fe]\footnote{[$\alpha$/Fe] = 1/4 ([Mg/Fe] + [Si/Fe] + [Ca/Fe] + [Ti/Fe])}, [Mg/Fe], and [Ti/Fe] ratios, measured at $-0.017 \pm 0.015$ dex, $-0.047 \pm 0.012$ dex, and $-0.012 \pm 0.010$ dex, respectively, which align with thin-disc kinematics and are consistent with findings in \citet[Figure 1]{Adibekyan:2012A&A...545A..32A}, \citet[Figure 3]{Fuhrmann:2017MNRAS.464.2610F}, and \citet[Figure 23]{Bensby:2014A&A...562A..71B}.

Additionally, we performed $\chi^2$ fitting of the spectral energy distribution (SED) of HIP 8522 (see bottom panel of Figure \ref{fig:toomre_sed}) using archival photometry spanning from $\lambda = 0.355~\mu m$ to $\lambda = 22.08~\mu m$ within the Virtual Observatory SED Analyser (VOSA; \citealt{2008A&A...492..277B}). The data includes observations from the Javalambre Physics of the Accelerating Universe Astrophysical Survey (J-PAS; \citealt{2014arXiv1403.5237B}), the Javalambre-Photometric Local Universe Survey (J-PLUS; \citealt{2019A&A...622A.176C}), \emph{Gaia} DR3 \citep{2023A&A...674A...1G}, the Hubble Space Telescope (HST; \citealt{2011ApJS..193...27W}), the Wide-field Infrared Survey Explorer (WISE; \citealt{2010AJ....140.1868W}), the 2-Micron All-Sky Survey (2MASS; \citealt{Ochsenbein:2000A&AS..143...23O, skrutskie2006}), and the Japan Aerospace Exploration Agency (JAXA) AKARI telescope \citep{2007PASJ...59S.369M}. The nominal values obtained with the {\sc BT-Settl} models grid \citep{Allard:2011ASPC..448...91A} effective temperature ($5700$ K), and surface gravity ($4.5$ dex), and metallicity (0.0 dex) are consistent with those derived from high-resolution spectroscopy (see Table \ref{tab:fundamental parameters}).

To discard a hidden companion, we performed a two-component model fit to the observed SED, where the primary component was fixed using the best-fit parameters from the single-star model, representing the known star in the system. The secondary component was varied across a range of effective temperatures ($T_{\rm eff} = 400-2300~K$) and surface gravities ($\log(g) = 4.5-6$ dex) within {\sc BT-Settl} models grid consistent with potential ultra-cool star and brown dwarf companions \citep{2000nlod.book.....R}, and varying between $5000-80 000$ K and $\log(g) = 6.5-9.0$ dex within models grid for white dwarfs \citep{2010MmSAI..81..921K}. Nevertheless, the resulting fits did not significantly improve upon the single-star model, and no compelling evidence for a binary companion, either an ultra-cool object or a degenerate star, was found, which supports the conclusion that HIP 8522 is a single star. 

It is worth mentioning that Gaia detects a star (\textit{Gaia} DR3 297548019937754752) less than three arcseconds away from HIP 8522, and two others (\textit{Gaia} DR3 297548019937986048 and \textit{Gaia} DR3 297548019937754496) within 40 arcseconds. Only \textit{Gaia} DR3 297548019937754496 has astrometric data, but its parallax (0.089 mas) differs from HIP 8522's (20.651 mas), indicating they are not associated. No astrometric or velocity information is available for the other two, making it difficult to determine whether they are associated with HIP 8522. Nonetheless, it is possible that none of these stars is gravitationally bound to HIP 8522, as no significant radial velocity (RV) variations are found (see Section \ref{sec:rv}).

\begin{table*}
    %\scriptsize
    \centering
    \caption{Quantities measured in the SOPHIE spectra of HIP 8522 using the \texttt{sophie-toolkit} \citep{Martioli2023}.}
    \begin{tabular}{ccccccccccccc}
        \hline
        \hline
        Time & RV & $\sigma_{\rm RV}$ & Exp. & S/N & FWHM & $\sigma_{\rm FWHM}$ & Bis & $\sigma_{\rm Bis}$ & S$_{\rm MW}$ & $\sigma_{\rm S_{\rm MW}}$  & H$\alpha$ & $\sigma_{{\rm H}\alpha}$ \\
        BJD & km\,s$^{-1}$ & km\,s$^{-1}$ & s & @649\,nm & km\,s$^{-1}$ & km\,s$^{-1}$ & km\,s$^{-1}$ & km\,s$^{-1}$ &  &  &  &  \\
        \hline
        2454408.39528 & -1.9357 & 0.0006 & 300 & 70 & 8.201 & 0.028 & 0.014 & 0.003 & 0.461 & 0.064 & 0.3049 & 0.0011 \\
        2454496.30200 & -1.8948 & 0.0013 & 900 & 29 & 8.289 & 0.030 & 0.018 & 0.010 & 0.349 & 0.053 & 0.3206 & 0.0031 \\
        2454694.62711 & -1.9016 & 0.0007 & 362 & 53 & 8.227 & 0.028 & 0.013 & 0.004 & 0.487 & 0.067 & 0.3241 & 0.0010 \\
        2454720.58791 & -1.9070 & 0.0006 & 334 & 55 & 8.235 & 0.028 & 0.008 & 0.009 & 0.453 & 0.063 & 0.3115 & 0.0014 \\
        2454725.59654 & -1.9173 & 0.0007 & 347 & 54 & 8.193 & 0.029 & 0.010 & 0.006 & 0.477 & 0.066 & 0.3113 & 0.0009 \\
        2454726.54247 & -1.9152 & 0.0006 & 269 & 54 & 8.192 & 0.028 & 0.006 & 0.004 & 0.440 & 0.062 & 0.3138 & 0.0010 \\
        2454729.60029 & -1.9197 & 0.0006 & 306 & 54 & 8.230 & 0.029 & 0.028 & 0.004 & 0.512 & 0.069 & 0.3222 & 0.0012 \\
        2457404.31316 & -1.9639 & 0.0019 & 300 & 15 & 8.255 & 0.033 & -0.036 & 0.027 & 0.122 & 0.035 &0.3310 & 0.0061 \\        
        \hline
        \hline
    \end{tabular}
    \label{tab:sophiequantities}
\end{table*}

\begin{figure}
\includegraphics[width=1\columnwidth]{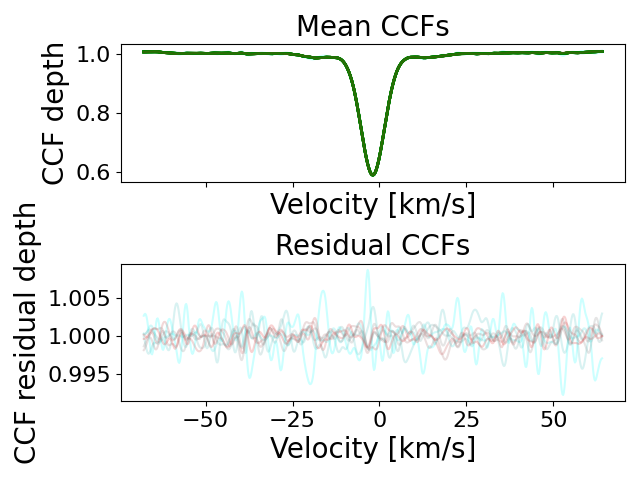}
 \caption{SOPHIE CCFs of HIP 8522. The top panel shows the mean CCF obtained for the SOPHIE spectra. The lower panel shows the individual CCFs with the mean CCF subtracted, where the color code shows the first epoch in dark blue and the last epoch in dark red.}
 \label{fig:sophieccfs}
\end{figure}

\begin{figure*}
\includegraphics[width=2.1\columnwidth]{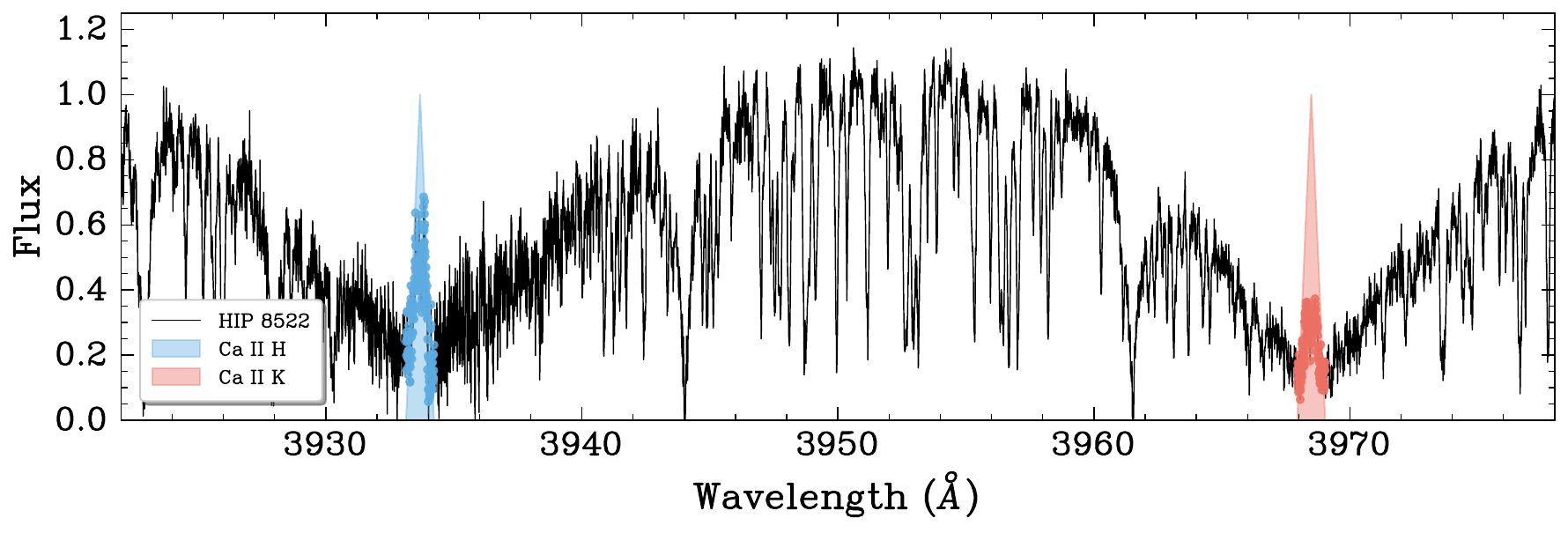}
\centering
 \caption{Measurement of the S-index in the SOPHIE template spectrum (solid black line) around the Ca II H and K line regions, including the integration windows for the line cores (blue and red shaded triangles). The cores of the Ca H (blue) and K (red) lines are pronounced, providing evidence of strong chromospheric activity in HIP 8522.}
 \label{fig:sophiesindex}
\end{figure*}

\begin{figure}
 $\begin{array}{c}
 \includegraphics[width=\columnwidth]{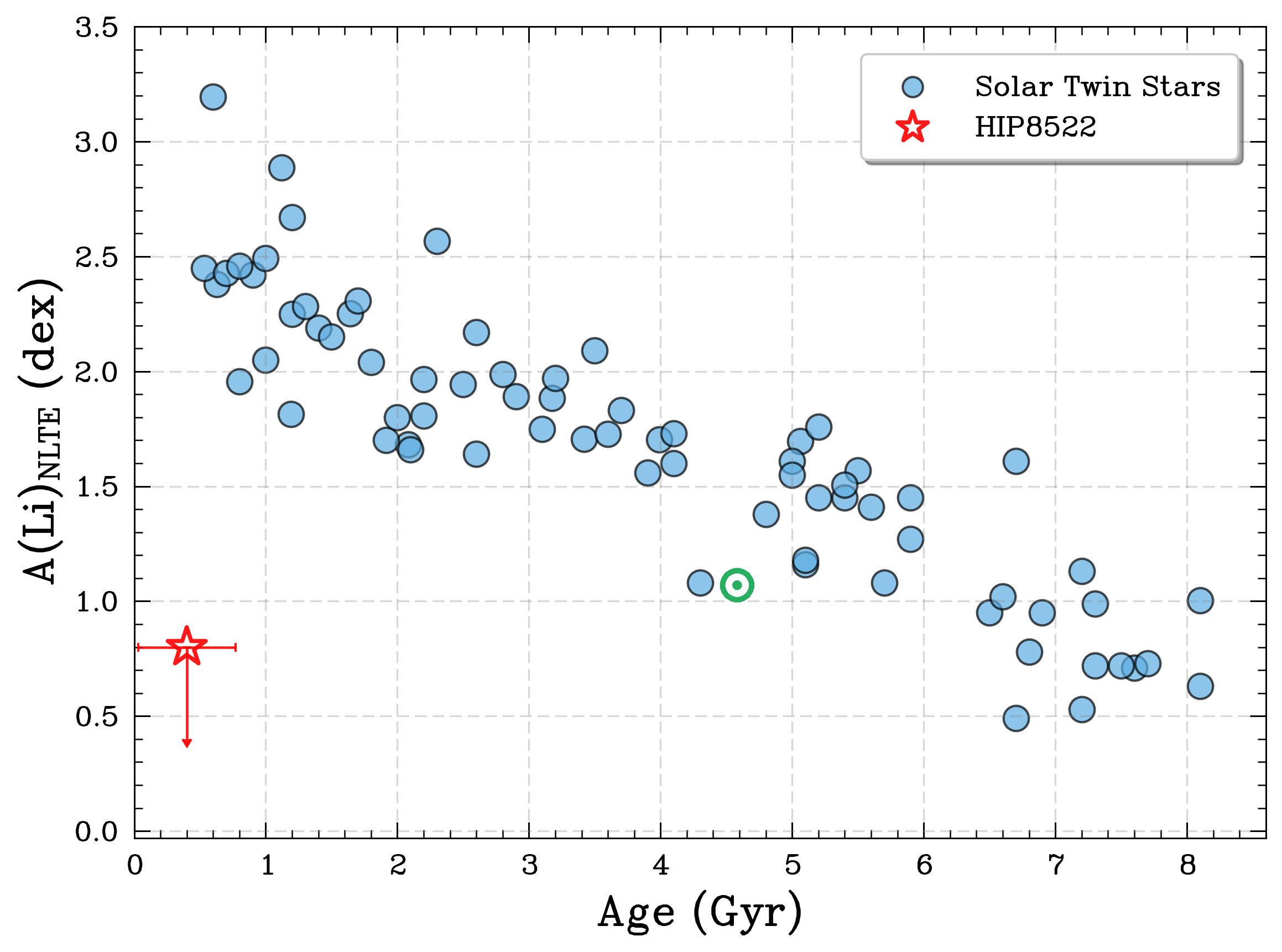} \\
 \includegraphics[width=\columnwidth]{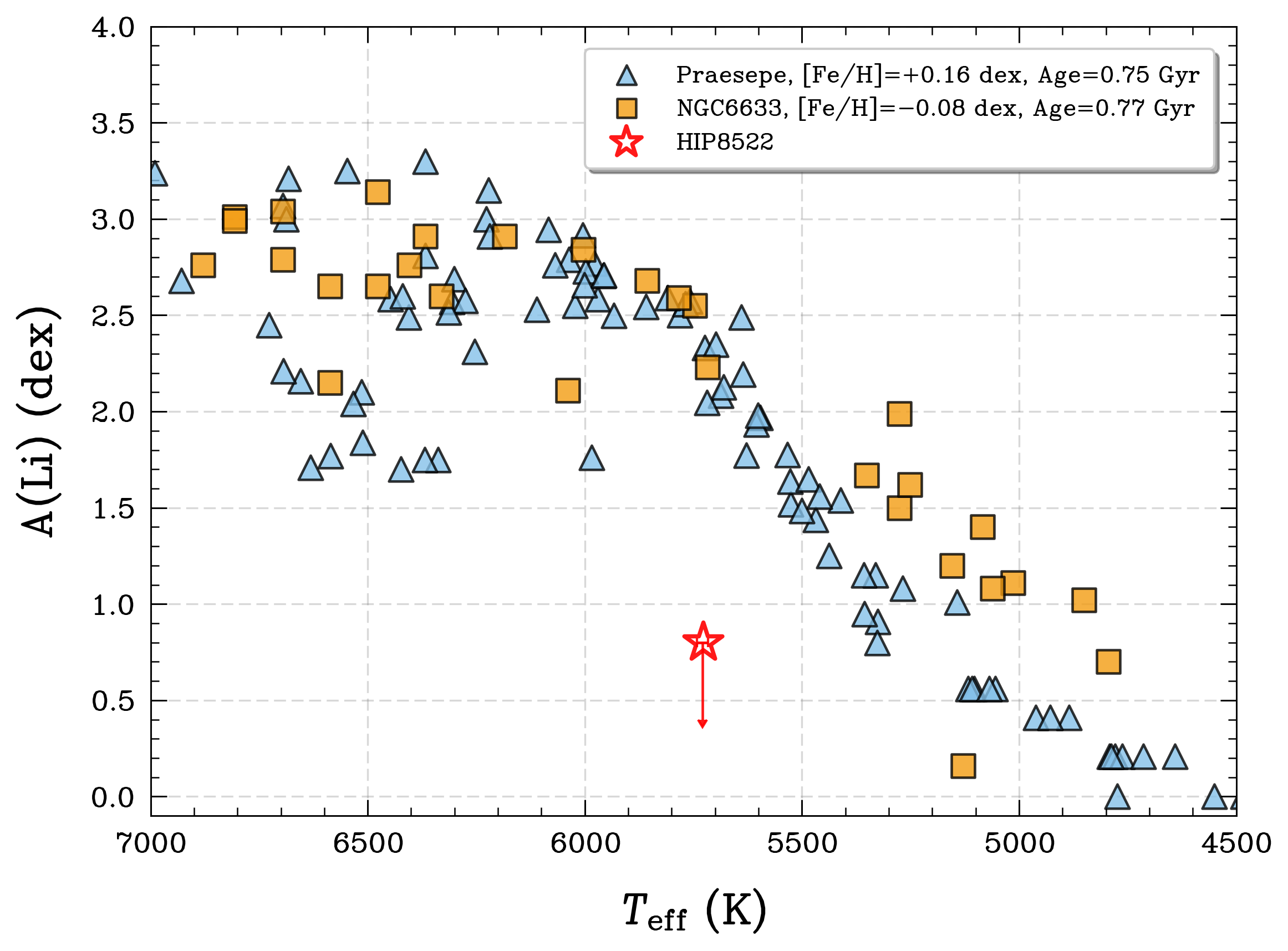} \\
 \includegraphics[width=\columnwidth]{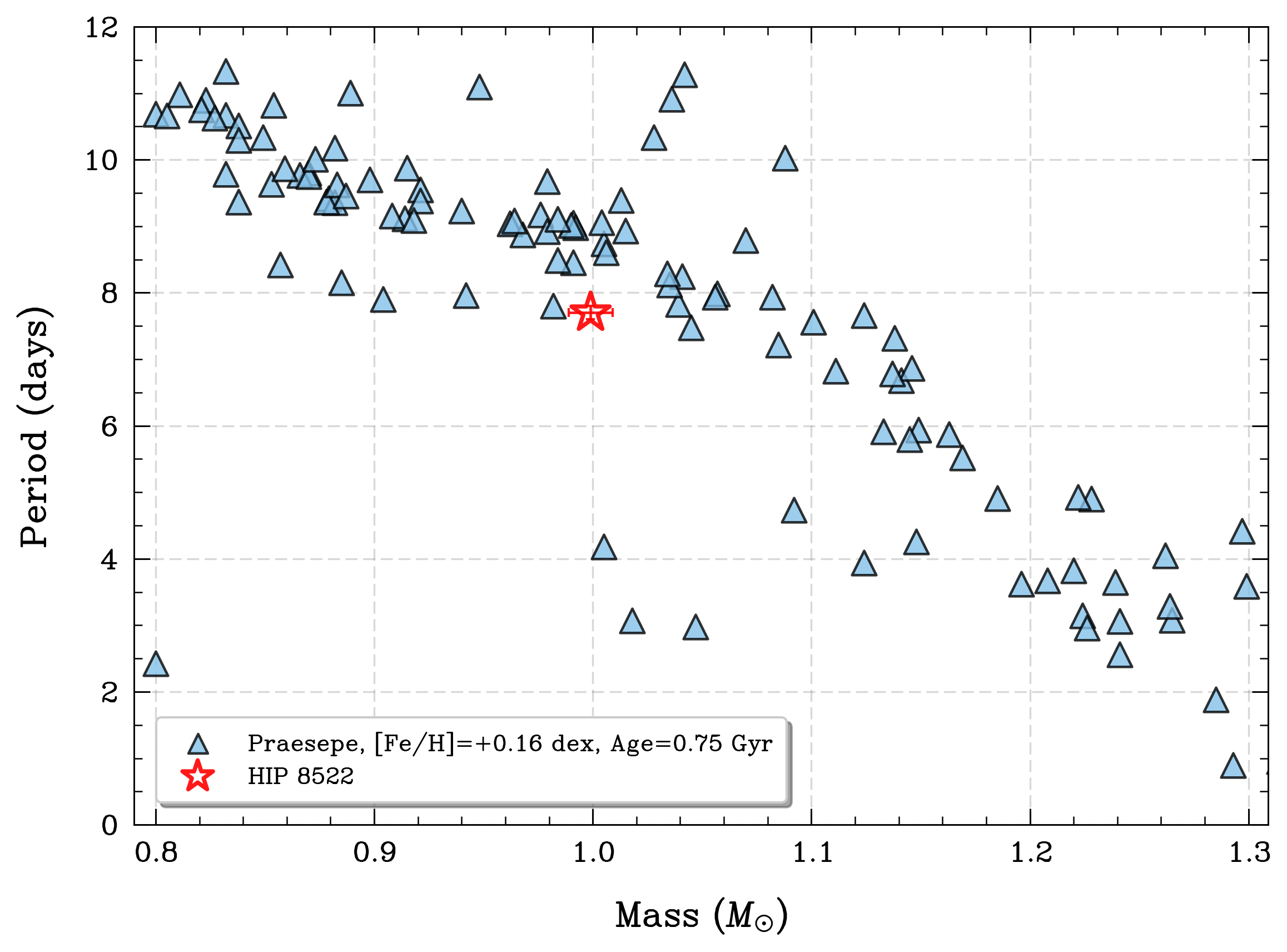}
 \end{array}$
 \centering
 \caption{ \textbf{Upper panel:} NLTE lithium abundance as a function of age for solar twins from the Inti survey \citep{Yana_Galarza:2021MNRAS.504.1873Y}. The Sun is plotted as a green solar standard symbol. \textbf{Middle panel:} Lithium evolution for the young open clusters Praesepe (triangles; age of $0.75$ Gyr) and NGC 6633 (squares; $0.77$ Gyr) as a function of the star's effective temperature. Data are from \citet{Jeffries:1997MNRAS.292..177J} and \citet{Jeffries:2002MNRAS.336.1109J}. \textbf{Bottom panel:} Rotational period as a function of mass for Praesepe (triangles). Data are from \citet{Godoy:2021ApJS..257...46G, Agueros:2011ApJ...740..110A} and \citet{Douglas:2017ApJ...842...83D, Douglas:2019ApJ...879..100D}. The red star represents HIP 8522 in all the panels.}
 \label{fig:unique}
\end{figure}

\begin{figure*}
\includegraphics[width=2.1\columnwidth]{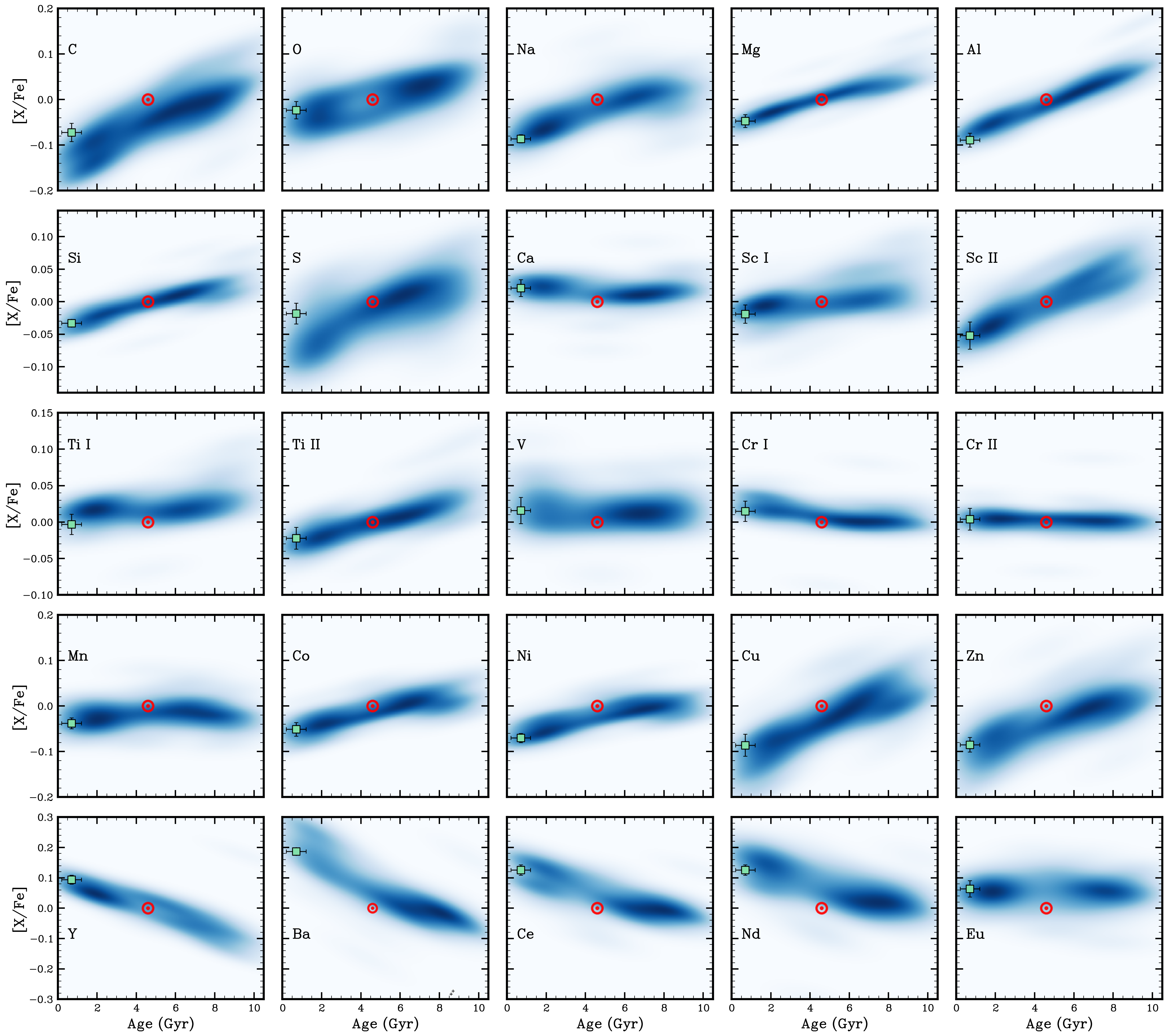}
 \centering
 \caption{Density plot of elemental abundances for solar analogs \citep{Bedell:2018ApJ...865...68B, Spina:2018MNRAS.474.2580S, Yana_Galarza:2021MNRAS.504.1873Y} in the thin-disc, as a function of age. The green square symbols represent HIP 8522, while the Sun is plotted as a red solar standard symbol. The ages of the solar analogs were estimated using isochrone fitting (see Section \ref{sec:ages}), the same method applied to HIP 8522.}
 \label{fig:GCE_evolution}
\end{figure*}

\section{Radial Velocities and Activity Indices from the SOPHIE Spectra} \label{sec:rv}

Among the spectroscopic observations presented in this paper, the SOPHIE data has the highest long-term stability, allowing us to investigate potential radial velocity variations due to the presence of stellar or sub-stellar companions. We analyzed the SOPHIE spectra using the \textsc{sophie-toolkit} \citep{Martioli2023}, which computes the cross-correlation function (CCF; see Fig. \ref{fig:sophieccfs}), radial velocities (RV), a stacked template spectrum, and activity indices such as the full width at half maximum (FWHM), bisector span (BIS), S-index, and H$\alpha$. Table \ref{tab:sophiequantities} presents these quantities.  

We measured a mean RV of $-1.919\pm0.020$~km\,s$^{-1}$ for all SOPHIE spectra of HIP 8522. When considering only the six spectra obtained in 2008, the mean RV is $-1.909\pm0.009$~km\,s$^{-1}$, showing a dispersion of only 9~m\,s$^{-1}$. These RV measurements do not cover a sufficient time baseline and cadence to detect signals of small amplitudes; however, they are adequate to rule out the existence of any massive close-in companions. For instance, a hypothetical sub-stellar object with a mass greater than 13~M$_{\rm Jup}$ would cause an RV variation with a semi-amplitude greater than $\sim500$~m\,s$^{-1}$. A semi-amplitude of 20~m\,s$^{-1}$, comparable to the RV dispersion of our data, would correspond to a companion of 0.5~M$_{\rm Jup}$, assuming a circular orbit. Moreover, we calculated a correlation coefficient of 0.72 between the RVs and both the H$\alpha$ index and the CCF FWHM. This indicates that the observed RV variability likely originates from asymmetries in the line profiles caused by stellar activity. Therefore, the mass limit of 0.5~M$_{\rm Jup}$ for a sub-stellar companion to HIP 8522 could be significantly lower.

Table \ref{tab:sophiequantities} presents the measurements of some activity indices for HIP 8522, obtained from the SOPHIE spectra. We have also calculated the same quantities for the SOPHIE stacked template spectrum.  For instance, we obtained the H$\alpha = 0.290\pm0.003$, and the S-index as illustrated in Fig. \ref{fig:sophiesindex}. The S-index has been calibrated to the Mt. Wilson Observatory (MWO) system \citep[$S_{\rm MW}$,][]{Wilson1968,Egeland2017} following the calibration recipe of \cite{Martioli2023}, which gives S$_{\rm MW} = 0.402\pm0.01$. We convert the Mount-Wilson S-index to the chromospheric activity index $\log{\rm R'}_{\rm HK}$ using the \cite{Czesla2019} recipe that includes the photospheric correction \citep[e.g.,][]{Mittag2013}. The result was $\log{\rm R'}_{\rm HK}(B-V)=-4.26\pm0.08$. This level of chromospheric activity is consistent with a very young age.

\section{Discussion} \label{sec:disc}

\subsection{HIP 8522: A Remarkably Unique Solar Twin}
HIP 8522 is a young solar twin with the lowest observed lithium content so far, making it unique among other stars. The upper panel of Figure \ref{fig:unique} shows the evolution of lithium in solar twins (blue circles) over time, where lithium is depleted by an additional mixing mechanism that transports lithium from the surface to below the convective zone, where the temperature is approximately $2.5 \times 10^{6}$ K -- sufficient to destroy lithium. For more details we refer to \citet{Charbonnel:2005Sci...309.2189C, Carlos:2016A&A...587A.100C, Castro:2016A&A...590A..94C, Baraffe:2017A&A...597A..19B, Martos:2023MNRAS.522.3217M}. HIP 8522 (red star) clearly does not follow the lithium evolution of solar twins. On the contrary, it seems that a highly efficient mixing mechanism allowed lithium to be destroyed in its very early stages. Typical young solar twins (age $<$ 1 Gyr) are rich in lithium, with levels between 2.5 and 3.3 dex (see upper panel of Fig. \ref{fig:unique}). This indicates that HIP 8522 is lithium-deficient by approximately 1.7 to 2.5 dex, or about 50 to 300 times lower.

Following the evolution of the lithium trend in solar twins, the lithium problem would be resolved if our star were approximately 8 Gyr old, as reported by \citet{Notsu:2017PASJ...69...12N}. This would cast doubt on the isochronal age calculation; however, we have confirmed HIP 8522's age through chemical age clocks, as well as activity and rotation age correlations (see discussions in Section \ref{sec:ages}), which serve as independent age estimators. Furthermore, when comparing lithium with effective temperature rather than with isochronal age, HIP 8522 remains far below the lithium abundance trend of stars in open clusters of similar age (see middle panel of Figure \ref{fig:unique}), confirming that the rarity of our star is not due to an incorrect estimation of its age. We explored whether other elements display anomalies similar to lithium. Figure \ref{fig:GCE_evolution} shows that the elemental abundances of HIP 8522 (green squares) follow the chemical evolution of solar analogs in the Galactic thin-disc, indicating that no anomalies are found in elements other than lithium.

\begin{figure*}
\includegraphics[width=2.1\columnwidth]{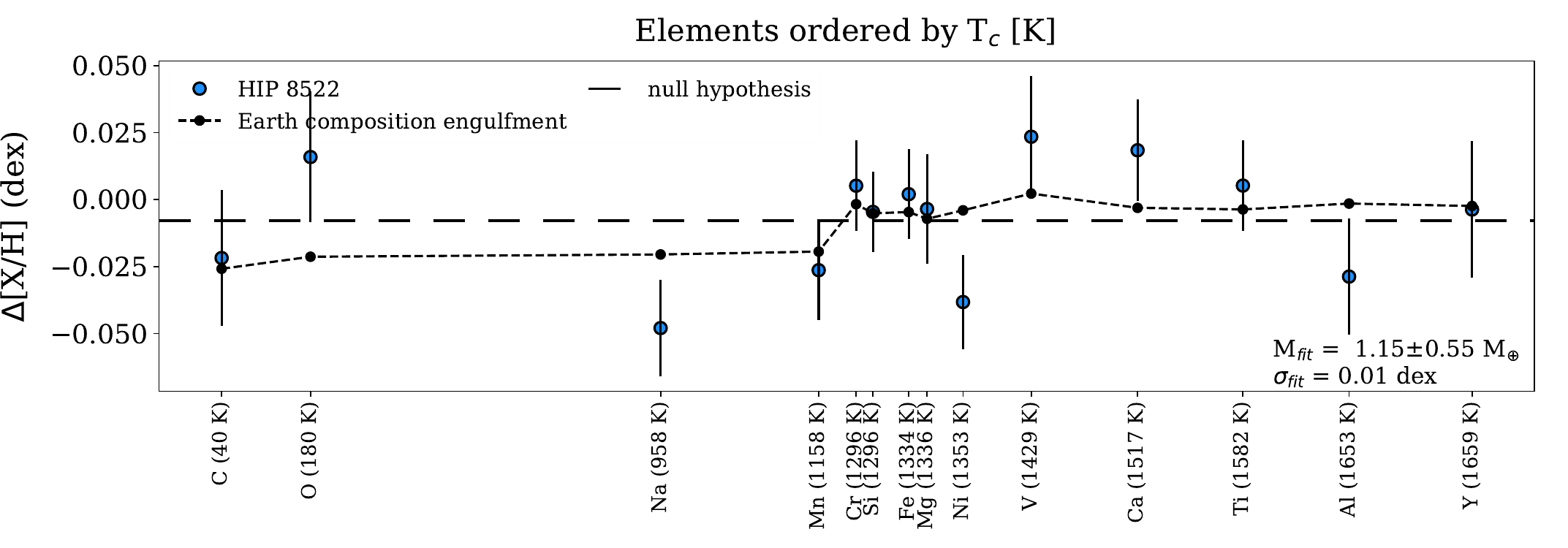}
 \centering
 \caption{Observed GCE-corrected differential abundance (HIP 8522 - Sun; blue circles) as a function of elements ordered by condensation temperature. The black circles represent the simulated planetary engulfment of 1.15 $M_{\oplus}$ of Earth bulk composition. The long-dash line indicates the limit of engulfment for the null hypothesis.}
 \label{fig:accretion_abs}
\end{figure*}

The lithium content of HIP 8522 differs from that of solar twins of similar age and also shows differences when compared to stars in open clusters at the same evolutionary stage. The middle panel of Figure \ref{fig:unique} shows the lithium evolution in two open clusters of similar age ($\sim$0.8 Gyr) but with different ranges of metallicity, encompassing HIP 8522's metallicity. HIP 8522 is clearly below the lithium trend. However, its rotational period ($\sim$7 days) appears to be typical in the period versus age-mass relation (see the bottom panel of Figure \ref{fig:unique}), confirming the uniqueness of HIP 8522 among young stars.  

\citet{Bouvier:2016A&A...590A..78B} reported a low-lithium star (Mon 1115, EW = 75 m\AA, \teff\ = 4833 K) in the young cluster NGC 2264 ($\sim$0.5 Gyr). However, they excluded it from their analysis, suggesting further investigation was needed to confirm its membership. The EW of HIP 8522 is calculated to be $3.64\pm2.10$ m\AA, representing a rarity compared to stars in NGC 2264 (see Bouvier's Figure 1). If HIP 8522 is not the first unique star discovery, it is at least the first among field stars. Additionally, HIP 8522 offers a unique opportunity to study stellar interiors. Its brightness ($G = 8.4$ mag) enables further studies with high resolution, high S/N, and broader spectral coverage, including exploring Beryllium and determining its activity cycle. The following subsection will outline potential scenarios that could explain the observed low lithium content in HIP 8522.

\subsection{Planet Engulfment Hypothesis}
\label{sub:engulfment}
When a solar twin is reported as chemically anomalous, it refers to the level of refractory elements relative to the Sun. A star that shows a similar level or deficiency in these elements (negative slope in the condensation temperature plane) is believed to potentially have its refractory elements sequestered by rocky exoplanets \citep[e.g.,][]{Melendez:2009ApJ...704L..66M, Ramirez:2010A&A...521A..33R, Yana:2021MNRAS.502L.104Y}. On the contrary, when a star displays an enhancement of refractory elements (positive slope in the \Tc\ plane), it is usually attributed to planet engulfment events \citep[e.g.,][]{Melendez:2017A&A...597A..34M, Yana_Galarza:2021ApJ...922..129G}. As HIP 8522 shows a moderate enhancement in refractory elements, even after correcting for GCE (see Figure \ref{fig:tc_abundances}), it is reasonable to consider that a planet engulfment might have occurred during the early stages of this star. We tested this possibility using the planet engulfment model from \cite{Behmard:2023MNRAS.521.2969B}, which allows tracking the evolution of each individual abundance over time following an engulfment at ZAMS. Figure \ref{fig:accretion_abs} shows that the engulfment of 1.15 $\pm$ 0.55 $M_{\oplus}$ of bulk Earth composition is necessary to nearly reproduce the observed GCE-corrected abundances. However, the Bayesian evidence difference, $\Delta$ln$(Z)$\footnote{Ratio between the engulfment model and a comparison flat model to assess the significance of the engulfment trend. For more details, we refer to \citet{Behmard:2023MNRAS.521.2969B}.}, estimated to be 1.43, is statistically insufficient to confirm the engulfment hypothesis on solid grounds. This means that while the engulfment model is preferred over the flat model (long-dash lines in Figure \ref{fig:accretion_abs}), the evidence is not strong enough for a definitive conclusion. It is worth mentioning that the $\Delta$ln$(Z)$ analysis is unable to detect signatures of planetary engulfment for $M < 10 M_{\oplus}$, as indicated by \cite{Behmard:2023MNRAS.521.2969B}.

Nevertheless, how does lithium evolve during planet engulfment? It is believed that during this process, not only are refractory elements significantly enhanced, but lithium is as well. This has been observed in some components of twin binary systems \citep[e.g.,][]{Saffe:2017A&A...604L...4S, Yana_Galarza:2021ApJ...922..129G, Spina:2021NatAs...5.1163S}. However, others display significant enhancement in refractory elements but not in lithium \citep{Flores:2024MNRAS.52710016F, Miquelarena:2024A&A...688A..73M, Yana:2024ApJ...974..122G}. \citet{Behmard:2023MNRAS.521.2969B} suggested that the signatures of planet engulfment depend on various factors such as stellar mass, convective zone size, and the timing of accretion, so the signatures of planet engulfment in some binary systems might be erased by the time they are observed. \citet{Sevilla:2022MNRAS.516.3354S} conducted detailed simulations of planet engulfment using {\sc MESA} stellar evolution models \citep{paxton2011, paxton2013, paxton2015, paxton2018, paxton2019} and demonstrated that the level and duration of lithium enrichment strongly depend on mixing mechanisms within the stellar interior. \citet{Theado:2012ApJ...744..123T} found similar results, concluding that planet engulfment could deplete lithium to below pre-engulfment levels.

\begin{figure}
 \includegraphics[width=\columnwidth]{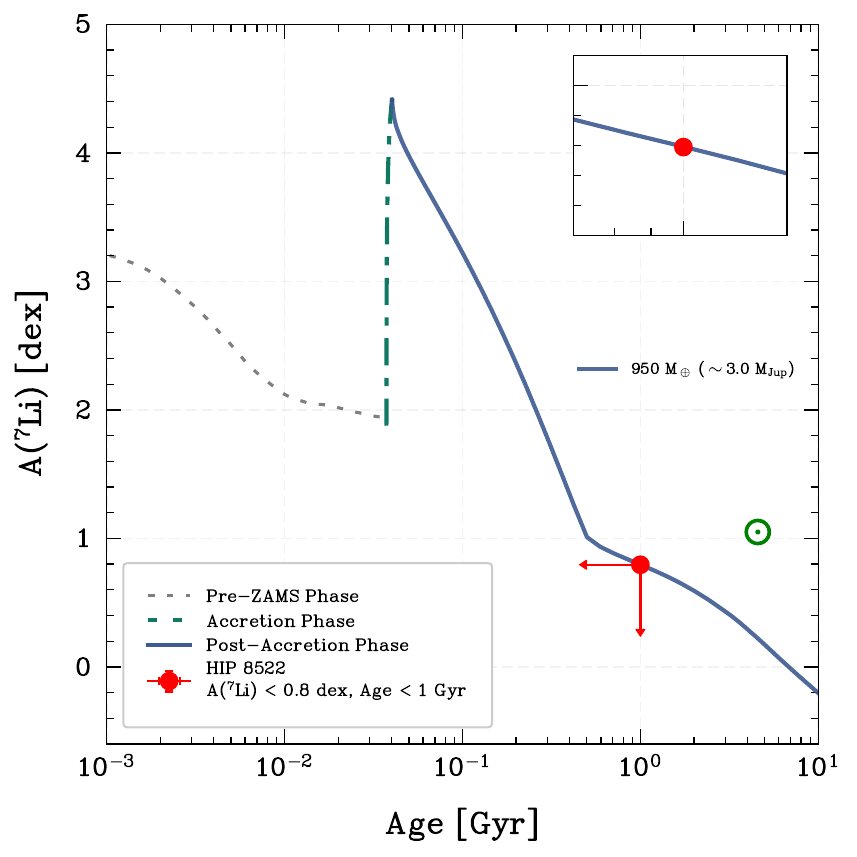}
 \centering
 \caption{Lithium abundance for HIP 8522 following a $\sim 950~M_{\oplus}$ ($\sim 3~M_{\rm Jup}$) engulfment model of bulk Earth composition as a function of time. The three stages of the simulation are shown as dashed grey lines (Pre-ZAMS), green dashed-dotted lines (planetary engulfment at ZAMS), and a solid blue line (post-accretion). HIP 8522 is represented by the red circle, while the Sun is represented by the green standard symbol.}
 \label{fig:accretion_Li}
\end{figure}

Starting from the principle that lithium is rapidly destroyed within a few Myr after engulfment through mechanisms such as atomic diffusion, overshoot mixing, and thermohaline mixing, we used the {\sc MESA} {\tt inlists} provided by \citet{Sevilla:2022MNRAS.516.3354S} to model planet engulfment and reproduce the lithium observed in HIP 8522. The model runs in three stages. The first stage evolves the star up to its ZAMS phase with a metallicity value of $Z = 0.017$ dex, close to that of the Sun, yielding a \li\ $\sim$ 3.3 dex. The only mixing process during this phase is convection. The second stage simulates planetary engulfment via the accretion of bulk Earth composition material \citep{MCDONOUGH1995223} at a rate of $3\times 10^{-4} M_{\oplus}$ yr$^{-1}$ at ZAMS. The final stage evolves the star until the end of its main sequence lifetime. Phases 2 and 3 include mixing mechanisms such as convective overshoot mixing, thermohaline mixing, and atomic diffusion. None of the stages include rotation. For more details in the {\tt inlists}'s set up, we refer to \citet{Sevilla:2022MNRAS.516.3354S}. Figure \ref{fig:accretion_Li} shows the evolution of lithium over time before, during, and after the accretion of $\sim$950 $M_{\oplus}$ of bulk Earth composition, which is the mass required to explain the observed lithium in HIP 8522. However, during planet engulfment, refractory elements must also be enhanced, and the signatures of such a significant engulfed mass should still be observable. For instance our simulation predicts an iron abundance of \feh $\sim0.4$ dex, which is inconsistent with the observed metallicity (\feh\ = 0.005 dex) of HIP 8522. The other elements show a similar level of enhancement. The planetary engulfment scenario seems unlikely due to the inconsistency between the observed and predicted abundances, aside from lithium.

\subsection{Blue Straggler}
Blue stragglers (BSSs) are mainly found in open clusters. Although they belong to the same population as the other stars in the cluster, they appear anomalously younger and hotter. Their formation is commonly attributed to mechanisms like merging with other stars or accreting mass from a companion, which makes them exceptionally bright and blue \citep[e.g.,][]{Sandage:1953AJ.....58...61S, 1989AJ.....98..217L, 1990AJ....100..469M, Sills:2000ApJ...535..298S, Ryan:2001ApJ...547..231R, 2019MNRAS.486.1220F}. Another key feature of BSs is their high lithium depletion, a trait also observed in HIP 8522. \citet{Schirbel:2015A&A...584A.116S} reported the discovery of HIP 10725, a solar analog with unusually low lithium ($<0.9$ dex) and beryllium ($<0.2$ dex) levels for its age ($\sim$5.2 Gyr). They concluded that this star is a field blue straggler, likely formed by accreting mass from a companion. This conclusion is based on the following findings: \textit{i)} HIP 10725 shows an unusually high excess of $s$-process elements, likely due to pollution from an Asymptotic Giant Branch (AGB) star; \textit{ii)} variations in radial velocity, up to 70 km s$^{-1}$, suggest the presence of a companion, probably a white dwarf; and \textit{iii)} the high projected rotational velocity (\vsini\ $= 3.3$ km s$^{-1}$) and chromospheric activity levels ($R_{\rm{HK}} = -4.51$) are inconsistent with the isochronal age estimated at $\sim 5.2^{+1.9}_{-2.1}$ Gyr.

\begin{figure}
 \includegraphics[width=\columnwidth]{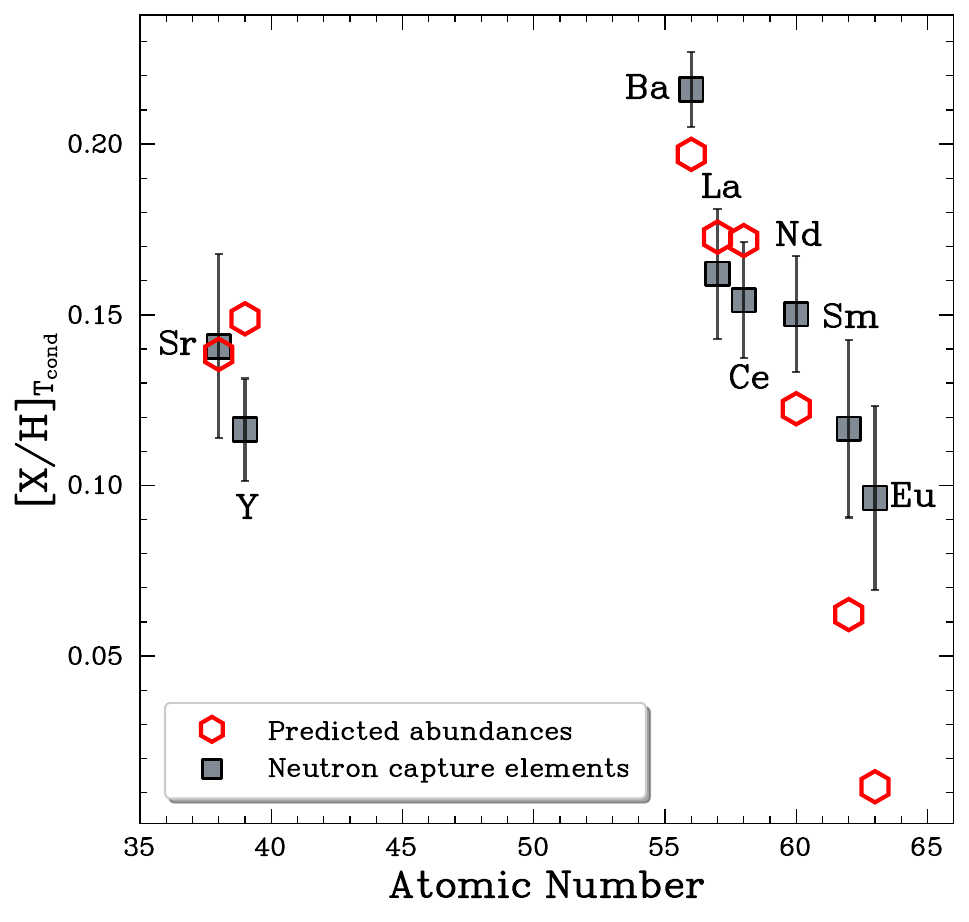}
 \centering
 \caption{Comparison of abundances corrected for condensation temperature effects, [X/H]$_{\rm{T_{cond}}}$ (gray squares), with predicted abundances (red open hexagons) from 1 \sm\ AGB pollution and 0.85\% dilution.}
 \label{fig:agb}
\end{figure}

Unlike HIP 10725, HIP 8522 is younger and exhibits activity and rotation levels typical of young solar twins ($<$1 Gyr), which align well with its isochronal age estimated of less than 1 Gyr. Among all the findings (\textit{i, ii, and iii}) listed above from \citet{Schirbel:2015A&A...584A.116S}, HIP 8522 appears consistent with the first, suggesting that the excess of $s$-process elements is likely due to pollution from an AGB star. Indeed, the upper panel of Fig. \ref{fig:tc_abundances} shows an excess of neutron capture elements (gray squares). To test this possibility, we followed the procedure given in and \citet{Melendez:2014ApJ...791...14M} and \citet{Yana_Galarza:2016A&A...589A..17Y}, which involves fitting elements with $Z \leq 30$ to remove the trend with condensation temperature from neutron-capture elements, using their Equation (2). To calculate the amount of pollution in the protocloud of HIP 8522 by an AGB star, we used models from \citet{Karakas:2010ApJ...713..374K}, which include diluted yields from a small fraction of AGB ejecta into a protocloud of solar composition and 1 \sm. We estimate a dilution of 0.85\% mass of AGB material to match the average observed enhancements in the $s$-process elements (Sr, Y, Ba, La, Ce, and Nd), as shown in Figure \ref{fig:agb}. However, an excess of $s$-process elements in HIP 8522 does not necessarily mean that AGB pollution destroyed lithium, as it likely did in HIP 10725. For example, young stars in clusters are enhanced in both $s$- and $r$-process elements \citep[e.g.,][]{DOrazi:2009ApJ...693L..31D, Maiorca:2011ApJ...736..120M, DOrazi:2012MNRAS.423.2789D} and do not show significant depletion of lithium. HIP 8522 is also enhanced in $r$-process elements, as reflected by Sm and Eu. Enhancements in neutron-capture elements have been observed in young solar twins, as shown by \citet{Cowley:2022MNRAS.512.3684C} in their Figure 2. Additionally, we did not detect any anomalies in the neutron-capture elements of HIP 8522 compared to other solar twins (see Fig, \ref{fig:GCE_evolution}). BSSs formed through mass transfer have close companions, which are expected to be white dwarfs with masses about 0.6 \sm\ \citep[][]{Geller:2011Natur.478..356G, Geller:2012AJ....144...54G}. Notwithstanding, the radial velocities of HIP 8522 do not show evidence of such a companion. On the contrary, if a companion exists, its mass should be lower than 0.5 M$_{\rm{Jup}}$ (see Section \ref{sec:rv}). Carbon and oxygen depletion are strong indicators of binary mass transfer \citep[see subsection 3.5 of][]{Wang:2024arXiv241010314W}, but these elements in HIP 8522 are typical of young solar twins (see Fig. \ref{fig:GCE_evolution}). Therefore, we conclude that a field blue straggler binary explanation for lithium deficiency is untenable.

Another mechanism for blue straggler formation involves binary mergers and stellar collisions. Nevertheless, the latter is more common in high stellar density environments, such as globular clusters and the cores of open clusters \citep[e.g.,][]{Hills:1976ApL....17...87H, Lombardi:2002ApJ...568..939L, Sills:2010MNRAS.407..277S}. If HIP 8522 is a field blue straggler, the stellar collision mechanism is unlikely, as discussed in \citet{Perets:2015ASSL..413..251P}. On the contrary, stellar mergers dominate in less crowded environments, such as Galactic fields and the outskirts of open clusters. Mergers occur when two close binary main-sequence stars come into contact or unstable mass transfer causes them to merge, eventually forming a more massive star \citep[e.g.,][]{Ferraro:2020RLSFN..31...19F, Wang:2024arXiv241010314W}. Three-dimensional magnetohydrodynamic (MHD) simulations of binary mergers support this hypothesis \citep{Schneider:2019Natur.574..211S}. However, these simulations involve massive stars of $\sim 8$ \sm, and no simulations exist yet for low-mass stars. Simulations have also demonstrated that BSSs do not retain clear memories of their origins and may have properties and surface chemical compositions similar to those of typical main-sequence single stars \citep{Glebbeek:2013MNRAS.434.3497G}, except for lithium, which is highly depleted \citep{Preston:2015ASSL..413...65P}. This depletion is caused by mixing processes resulting from the binary merger. Examples of such mixing include rotation-driven circulation \citep{Pinsonneault:1999ApJ...527..180P, Pinsonneault:2002ApJ...574..398P} and rotation due to angular momentum transfer \citep{Ryan:2002ApJ...571..501R}. 

We consider the possibility that HIP 8522 might be the progeny of a field blue straggler star, potentially resulting from a close binary merger that had previously depleted its surface lithium to very low levels but with its other parameters and composition similar to those of typical young solar twins. 

\subsection{Sub-stellar engulfment}
\citet{Cabezon:2023A&A...670A.155C} proposed an alternative scenario to the non-standard models, suggesting that interaction with sub-stellar objects could be the cause of the lithium depletion observed in solar-type stars. They performed 3D hydrodynamic simulations to explore how the accretion of a brown dwarf (BD; 0.01-0.019 \sm) by a solar-like main-sequence star (1 \sm\ and solar composition) could significantly alter the inner structure of the star and affect the Li content on its surface. The simulation explores the dynamics of the interaction, presenting three different scenarios: two collisions (head-on and grazing) and one merger. They found that in all scenarios, the companion is diluted into the star, and lithium is enhanced in the envelope of the MS star, with the enhancement being most significant in the case of the merger. A similar result was found by \citet{Sandquist:1998ApJ...506L..65S, Sandquist:2002ApJ...572.1012S}, who also used 3D hydrodynamical studies of the accretion of a sub-stellar object by a main sequence (MS) star. Therefore, given that simulations of brown dwarf accretion show an enhancement of lithium in 1 \sm~stars, which is inconsistent with the depleted lithium observed in HIP 8522, we ruled out this possibility. 

\subsection{Episodic Accretion Onto Young Stars}
Episodic accretion has significant implications for the formation and evolution of low-mass stars. \citet{Kenyon:1995ApJS..101..117K} proposed episodic accretion as a potential explanation for the luminosity spread observed in embedded protostars \citep[see also][]{Kenyon:1990AJ.....99..869K, Kenyon:1994AJ....108..251K, Evans:2009ApJS..181..321E, Enoch:2009ApJ...692..973E, Kryukova:2012AJ....144...31K, Dunham:2012ApJ...755..157D}. This hypothesis suggests that, rather than steady and uniform accretion from the protostellar disc onto the central protostar, the process occurs mainly in short but intense episodes, during which $0.01-0.1$ \sm\ can be accreted at rates of \.{M} $\geq 10^{-4}$ \sm\ yr$^{-1}$.

A luminosity spread is also observed in the Hertzsprung-Russell (HR) diagrams of young star-forming regions \citep{Hillenbrand:2009IAUS..258...81H}, though its origin remains under debate. \citet{Baraffe:2009ApJ...702L..27B, Baraffe:2010A&A...521A..44B} incorporated episodic accretion into protostar evolutionary models, reproducing the observed luminosity spread in young clusters. They also explored the effects of episodic accretion on lithium depletion using models with high accretion rates ($5\times 10^{-4} M_{\odot}$ yr$^{-1}$) onto very low-mass protostars with initial masses $>10$ $M_{\rm{Jup}}$. Their results suggest that the temperature required to develop a radiative core is proportional to the density. Both density and temperature influence the temperature at the bottom of the convective envelope, affecting the level of lithium depletion. Higher density results in higher temperatures, leading to more significant lithium depletion. For example, in episodic accretion sequences for an initial mass of 10 $M_{\rm{Jup}}$, the temperature at the base of the convective zone reaches $\sim 7\times 10^{6}$ K, causing complete lithium depletion at ages $<1$ Myr. Later, \citet{Baraffe:2017A&A...597A..19B} presented models of forming stars with more consistent accretion rates by coupling numerical hydrodynamics with stellar evolution models. They studied 60 models and found only one case that showed significant lithium depletion (see their Section 4.2). This model corresponds to a cold accretion scenario, which assumes that no accretion energy is absorbed by the protostar, similar to the model presented in \citet{Baraffe:2010A&A...521A..44B}. This finding aligns with HIP 8522, a rare example among the hundreds of reported solar twins, distinguished by its highly depleted lithium and potential evidence of an early episodic accretion history.

\section{Summary and Conclusions} \label{sec:conclusions}
We present the young solar twin HIP 8522, which shows the lowest lithium abundance detected among solar twins, suggesting it could either be a field blue straggler or the result of episodic early accretion. We determined its fundamental parameters (\teff, \logg, \feh, age, mass, radius) and chemical composition using high-resolution spectra from three different instruments (TS23, HDS, MIKE). Our findings suggest that HIP 8522 is young ($<$1 Gyr), supported by age-chemical clocks, stellar activity, and rotation-age relationships. The lithium abundance, estimated through spectral synthesis, provides an upper limit of $<$0.8 dex, which is unusually low for a young solar twin, as such stars typically have lithium levels between $2-3.3$ dex (see upper panel of Figure \ref{fig:unique}). The other chemical elements are typical compared to those of other young solar twins of similar age (see Figure \ref{fig:GCE_evolution}). HIP 8522's rotation and activity levels are also consistent with its age.

We explored the possibility of a sub-stellar or degenerate companion, but the RUWE (0.987), radial velocity, and the SED of the star provide no evidence. As the refractory elements show enhancement in the condensation temperature plane (see Figure \ref{fig:tc_abundances}), we investigate the possibility that planetary engulfment quickly depleted lithium in HIP 8522. We ran engulfment simulations with MESA and found that the accretion of 950 $M_{\oplus}$ ($\sim 3 M_{\rm{J}}$) can reproduce the observed lithium. Such massive accretion should significantly increase the levels of refractory elements, yet we do not observe this. Therefore, HIP 8522 likely did not undergo planet engulfment.

We explore the possibility that HIP 8522 is a blue straggler, potentially formed through mass accretion from a companion or the merging of a close binary. However, we do not find convincing evidence for mass transfer, as no companion has been detected via radial velocity, and the enhanced $s$-process elements observed are typical of young stars. On the other hand, the merging hypothesis appears plausible, as blue stragglers of this type tend to exhibit chemical properties similar to main-sequence stars, except for lithium, which is highly depleted. Another possibility is early episodic accretion, which suggests that HIP 8522 accretes material from the protostellar disc at high rates during short but intense episodes. This process increases its central temperature and density, potentially leading to temperatures at the bottom of the convective zone similar to those required for lithium burning. As a result, lithium may be quickly depleted within just a few Myr. The young solar twin HIP 8522 could represent the first observational evidence of early episodic accretion.

\section*{Acknowledgments}
%% Also note that the akcnowlodgment environment does not support long amounts of text. If you have a lot of people and institutions to acknowledge, do not use this command. Instead, create a new \section{Acknowledgments}.
%\begin{acknowledgments}
J.Y.G. acknowledges support from a Carnegie Fellowship. T.F. acknowledges support from Yale Graduate School of Arts and Sciences. H.R. acknowledges the support from NOIRLab, which is managed by the Association of Universities for Research in Astronomy (AURA) under a cooperative agreement with the National Science Foundation. D.L.O acknowledges the support from CNPq (PCI 301612/2024-2). E.M. acknowledges funding from FAPEMIG under project number APQ-02493-22 and a research productivity grant number 309829/2022-4 awarded by the CNPq, Brazil. R. L-V acknowledges support from CONAHCyT through a postdoctoral fellowship within the program ``Estancias Posdoctorales por M\'exico''.

The observations were carried out within the framework of Subaru-Keck/Subaru-Gemini time exchange program which is operated by the National Astronomical Observatory of Japan. We are honored and grateful for the opportunity of observing the Universe from Maunakea, which has the cultural, historical and natural significance in Hawaii.

This work has made use of data from the European Space Agency (ESA) mission Gaia (\url{https://www.cosmos.esa.int/gaia}), processed by the Gaia Data Processing and Analysis Consortium (DPAC, \url{https://www.cosmos.esa.int/web/gaia/dpac/consortium}). Funding for the DPAC has been provided by national institutions, in particular the institutions participating in the Gaia Multilateral Agreement. This publication makes use of VOSA, developed under the Spanish Virtual Observatory (\url{https://svo.cab.inta-csic.es}) project funded by MCIN/AEI/10.13039/501100011033/ through grant PID2020-112949GB-I00.
VOSA has been partially updated by using funding from the European Union's Horizon 2020 Research and Innovation Programme, under Grant Agreement nº 776403 (EXOPLANETS-A)

%\end{acknowledgments}


\vspace{5mm}
\facilities{Smith:TS23, Subaru:HDS, Magellan:Clay/MIKE, OHP:SOPHIE}


\software{
\textsc{numpy} \citep{van_der_Walt:2011CSE....13b..22V}, 
\textsc{matplotlib} \citep{Hunter:4160265}, 
\textsc{pandas} \citep{mckinney-proc-scipy-2010}, 
\textsc{astroquery} \citep{Ginsburg:2019AJ....157...98G}, 
\textsc{iraf} \citep{Tody:1986SPIE..627..733T}, 
\textsc{iSpec} \citep{Blanco:2014A&A...569A.111B, Blanco:2019MNRAS.486.2075B}, 
\textsc{Kapteyn} Package \citep{KapteynPackage}, 
\textsc{moog} \citep{Sneden:1973PhDT.......180S}, 
q$^{\textsc{2}}$ \citep{Ramirez:2014A&A...572A..48R}, 
\textsc{Gala} \citep{gala}}, 
\textsc{terra} \citep{Yana_Galarza:2016A&A...589A..65G}, 
\textsc{lightkkurve} \citep{2018ascl.soft12013L}, 
\textsc{astrobase} \citep{2021zndo...4445344B}, 
\textsc{VOSA} \citep{2008A&A...492..277B}. 



%\appendix


\bibliography{references}{}
\bibliographystyle{aasjournal}



\end{document}